\documentclass[floatfix,twocolumn]{aastex7}
\usepackage{amsmath}
\usepackage{bm}
\usepackage[normalem]{ulem}
\usepackage{wrapfig}
\usepackage{hyperref}
\usepackage{subfig}
\usepackage{xcolor}
\submitjournal{ApJ}

\shorttitle{ML inversions of RM measurements}
\shortauthors{Long et al.}

\graphicspath{{./}{figures/}}

\begin{document}


\title{\vspace{-8mm}A Convolutional Neural Network for the Recovery of Transfer Functions From Velocity-Resolved Reverberation Mapping Data}

\author{Kirk Long}
\affiliation{JILA, University of Colorado Boulder}
\email{kirk.long@colorado.edu}
\correspondingauthor{Kirk Long}
\email{kirk.long@colorado.edu}
\author{Keith Horne}
\affiliation{SUPA Physics and Astronomy, University of St Andrews}
\email{kdh1@st-andrews.ac.uk}
\author{Jason Dexter}
\affiliation{JILA, University of Colorado Boulder}
\email{jason.dexter@colorado.edu}
\author{Benoit Tremblay}
\affiliation{Environment and Climate Change Canada}
\email{benoit.tremblay.ec@gmail.com}

\begin{abstract}
One of the hallmarks of active galactic nuclei are that they are highly variable with time. In watching the spectra vary it has been observed that the emission-lines often appear to ``reverberate"\textemdash that is they vary in response to continuum variations assumed to originate close to the black hole. This critical observation underlies the reverberation mapping technique, an elegant physics experiment that has allowed us to characterize the environment around many supermassive black holes in nearby active galactic nuclei. Recent observations are of such quality that the response can be measured as a function of velocity across the emission-line, and in doing so we can construct velocity-delay maps that show the structure and physics of the gas in the broad-line region better than any other measurement to date. Unfortunately constructing such maps requires a deconvolution, and given that the data are often noisy and with gaps such deconvolutions are non-trivial. Here we present a novel deconvolution method for the recovery of velocity-delay maps using a custom convolutional neural network architecture, showcasing that such methods have great promise for the deconvolution of reverberation mapping data products. While we have designed this new method with the BLR in mind, in principle this technique could be applied to any reverberation deconvolution problem, including in the accretion disk and torus.
\end{abstract}

\section{Introduction}\label{sec:intro}
Variability is an observational hallmark of quasars and active galactic nuclei that has been exploited to infer many underlying physical processes in these systems that are otherwise not directly observable. Reverberation mapping is one such technique, where variations in one part of the spectrum are assumed to drive variations in another part\textemdash for example the AGN broad-lines are assumed to reprocess light from the continuum \citep{BLANDFORD_MCKEE_82}. The inner components of most AGN are too small to resolve visually with current instruments, but vary on scales short enough to allow for the reverberation mapping technique to provide insights into the structure and kinematics of regions where sources are unresolved, usually at broad-line region scales and smaller \citep{RM_Review_21,Peterson93_RM}. Even in sources whose BLRs are marginally resolved by advanced instruments such as GRAVITY \citep{GRAVITY19,GRAVITY+}, reverberation mapping data play a critical role in constraining what geometries and kinematics are possible for the BLR \citep{Long_2025,STORM_MODEL_SPECTRA,STORM_ECHO,STORM_CLOUDS} and may even enable the BLR to test the Hubble constant \citep{RM3c27319,Watson2011_RM_AGN_cosmological_distance}. 

In this paper we will focus on reverberation mapping as applied to the BLR, but note that the methods here are quite general and flexible and may be applied to other reverberation mapping problems. Originally reverberation mapping experiments in the BLR were 1D only \citep[i.e. see][for older reverberation results in NGC 5548]{KORISTA_95_NGC5548_RM, Krolik1991_NGC_5548,HorneWelshPeterson1991_EchoMapping_NGC5548}, that is they only measured the delay for the entire emission-line compared to the continuum. Advances in instrument and data processing techniques now allow for 2D reverberation mapping, where information can be obtained as a function of wavelength/velocity across the emission-line \citep{Welsh&Horne_EchoMaps, STORM_ECHO}. 

Formally, reverberation mapping assumes variations in emission from the gas in the broad-line region are related to variations in the continuum via a convolution:

\begin{equation}\label{variabilityEqn}
\Delta L(\nu,t) = \int_0^\infty \Psi(\nu,\tau)\Delta C(t-\tau)\mathrm{d}\tau = \Psi*\Delta C
\end{equation}
Here $\Psi(\nu,\tau)$ is the 2D \textit{frequency} and \textit{time} dependent \textit{response} function, which encodes how changes in the emission-line $\Delta L(\nu,t)$ ``respond" to changes in the continuum $\Delta C$. Thus nearly all of the interesting physics that govern the BLR are encoded within $\Psi$. 

If our data (and the system) were idealized and taken over long periods of time, one could simply solve for $\Psi$ from $\Delta L$ and $\Delta C$ in Equation \ref{variabilityEqn} using the convolution theorem:
\begin{align}\label{deconv}
\Psi &= \mathcal{F}^{-1}\left(\frac{\mathcal{F}(\Delta L)}{\mathcal{F}(\Delta C)}\right) = \mathcal{F}^{-1}\left(\frac{1}{\mathcal{F}(\Delta C)}\right)*\Delta L \\
&= \boxed{f(\Delta C)*\Delta L}
\end{align}
where $\mathcal{F}$ denotes the Fourier transform and $\mathcal{F}^{-1}$ its inverse \citep{BlandfordMckee1982_RM_Fourier}. In practice measurement problems prevent this, as the data are often sparse, noisy, and may not strictly follow a linear convolution at all times. This has motivated the development of many regularized inversion methods that attempt to control the noise-resolution trade-off and cope with noisy and gappy data to obtain the required inversions \citep{MEMEcho1,MEMEcho2,Krolik1995_regularizedLinearInversion}.

Here we describe a new method for solving this inversion problem using a Convolutional Neural Network (CNN). CNNs have long been successfully used for image deconvolution both in industry \citep{industry1_Li2021,industry2_DLGEO,MicrosoftLenovoDeconv} and astronomy \citep{Herbel_2018_ASTRO_PSFDL,Akhaury22_DL_GALDECONV}, but an analogous approach has yet to be applied to the deconvolution of BLR RM data. This approach is strongly physically and mathematically motivated. Decades of RM measurements have robustly shown that $\Delta C$,$\Delta L$, and $\Psi$ are connected, but the precise relationship is muddled by various physical and systematic sources of noise. As shown in the boxed form of Equation \ref{deconv}, an equivalent way of thinking about the recovery of $\Psi$ is that we are looking to find some complicated function of $\Delta C$ that we can convolve with $\Delta L$. 

A CNN is designed precisely with this task in mind: the network essentially ``learns" the potentially complex and highly nonlinear function $f(\Delta C)$ as the optimized weights are convolved with the input emission-line light curves to map back to the target response function. Thus the continuum and all of the errors both physical and systematic are absorbed into the parameters of the CNN, and in this approach what the CNN learns is what should be convolved with the input lightcurve to generate the correct transfer function. Because of the nature of the fundamental mathematical operation we are trying to approximate, we refer to this class of ML model as a (de)convolutional neural network ((D)CNN) throughout the rest of this paper.
We choose a CNN instead of a standard fully connected neural network because of the mathematics of the problem as described above, as standard fully connected
neural networks do not learn convolutional filters and thus struggle with spatial and temporal pattern recognition.
Another ML technique\textemdash variational autoencoders\textemdash has recently shown promise for recovering response functions for AGN coronae from X-ray lightcurves \citep{Deesamutara_AndyYoung_2025_ML_RM_XRAY}, further motivating an ML application to the BLR RM inversion problem, although the data quality is high enough in the BLR that
we choose the more direct CNN approach here.

In RM the relationship between line and continuum variations may well be more complicated than the linearized echo model in Equation \ref{variabilityEqn} \citep{Krolik1995_regularizedLinearInversion}. Additionally, traditional methods are computationally expensive, and while a (D)CNN is computationally expensive to train it is extremely cheap to use once trained. This strongly motivates the work described here, in which we use (D)CNNs to recover $\Psi$. Using a (D)CNN method allows for much more robust flexibility in obtaining $\Psi$ as in even our simplest models described here there are millions of parameters, but this flexibility comes at the cost of interpretability. The (D)CNN is somewhat of a black box, although we attempt to mitigate (largely unsuccessfully) this interpretability problem in Section \ref{sec:discussion}.

There are few, if any, interesting physics in the deconvolution itself, but failing to do it correctly prevents us from unraveling the true nature of the BLR. As the 2D lightcurves essentially represent a ``blurry" image, we follow a hybrid approach for image deconvolution inspired by \cite{MicrosoftLenovoDeconv} and \cite{RESNET_2015} by training a simple (D)CNN using Julia's \texttt{Flux.jl} framework \citep{flux1,flux2}, and in this paper we will showcase the promise of this method for the deconvolution of real RM data.

In the rest of this paper we explore the feasibility of using a custom-designed (D)CNN to solve this inversion problem. First, we describe our model architecture and training procedure in Section \ref{sec:models}. We showcase how these models perform on both 1D and 2D RM data in Sections \ref{sec:1D} and \ref{sec:2D}, respectively. We then describe how these models might be used on ``real'' problems through transfer learning in Section \ref{sec:transfer}. We include simple attempts to interpret how this black-box method works in Section \ref{sec:interpretability}. Finally, we offer concluding remarks and a discussion on the strengths and weaknesses of this approach, as well as future directions for this work in Section \ref{sec:discussion}.

\section{Summary of machine learning model architecture and implementation}
\label{sec:models}
\subsection{Model architecture}
We use Julia's \texttt{Flux.jl} framework for creating a custom CNN architecture. In our testing we find that it is possible to recover transfer functions of a variety of shapes with just a few simple layers. This architecture is summarized below, and a flowchart diagram of the architecture is shown in Figure \ref{fig:model} in the \hyperref[appendix]{Appendix}:
\begin{enumerate}
    \item We split the input into three independent paths, where each path has convolutional layers with filters corresponding to small, medium, and large velocity and temporal scales. This allows the model to learn information at each scale without being unduly influenced by the other filters. These scales are chosen arbitrarily, and here we have found using a small time scale filter size $k_{ts} = \rm{max}(2, \lfloor nT/100\rfloor)$, small velocity scale filter size $k_{vs} = \rm{max} (2, \lfloor nLC/100\rfloor)$, medium time scale filter size $k_{tm} = \rm{max}(5, \lfloor nT/50\rfloor)$, medium velocity scale filter size $k_{vm} = \rm{max} (5, \lfloor nLC/50\rfloor)$, large scale time scale filter size $k_{tl} = \rm{max}(7, \lfloor nT/5\rfloor)$, and large scale velocity filter size $k_{vl} = \rm{max} (7, \lfloor nLC/5\rfloor)$ works well. In the examples shown in this paper (both 1D and 2D) we always pass $1000$ elements in time, thus the filter sizes in time space are $k_{ts} = 10$, $k_{tm} = 20$, $k_{tl} = 200$. In our 2D examples we train models with both 25 and 50 possible velocity bins, thus these filter sizes are $k_{vs} = (2,2)$, $k_{vm} = (5,5)$, and $k_{vl} = (7, 10)$ respectively.
    \item The outputs from these independent paths produced after step 1 are then joined together and passed to a skip connection chain. Within the chain the model performs each convolutional filter in series going from small to large timescales and then small to large velocity scales, using the same filter sizes as described in step 1. At the end the results are added to the unadultered output of the split/join chain in the preceding layer. This design choice is inspired by RESNET \citep{RESNET_2015} and allows the model to recall previously learned information more easily, in this case information about each of the three possible scales we include independently. 
    \item To increase the depth of the neural network, we then perform the above sequence again, with the only difference being that the inputs to the split/join chain are now larger in the channel dimension as a result of the output from the skip connection chain. Memory limitations in computing the gradients limit this extra depth to our 1D models only, and in our 2D models we perform just a single pass of items 1 and 2.
    \item Finally, we apply a $(1,1)$ convolutional filter across the output from items 1-2 to output the model's prediction, which has the same size as the input dimensions in time and velocity space.
\end{enumerate}
After each convolutional layer we regularize by including batch normalization (BN) and dropout (Drop) layers and then apply a rectified linear unit (ReLU) activation function, with the exception of the final output layer which includes only the ReLU activation function after taking the final convolution. 

\subsection{Training procedure}
To train most of our models we use a synthetic AGN lightcurve generated according to a damped random walk \citep{DRW_2_Kozlowski2010, DRW_MacLeod2010} that is convolved with a few thousand possible model transfer functions to generate synthetic output lightcurves. We sample this generated lightcurve at a cadence that is $1/1000$th of the duration, i.e. equivalent to an observational campaign with a $1000$ day duration with the source observed daily. While 
damped random walks are likely an oversimplification of the true variability of quasars \citep{DRW_Complications_Kasliwal2015,Beyond_DRW_Yu2022}, they are fast and easy to compute. These synthetic output lightcurves are then subject to random white noise designed to simulate observational errors to first order, which we simulate by adding a random deviation from the truth drawn from a normal distribution with a width of $5\times10^{-3}$ the dynamic range of the lightcurve. We also randomly drop up to 10\% of the data points, a procedure meant to mimic bad data and/or observational gaps. These lightcurves are generated assuming that they have been properly de-trended, as there are potentially interesting physics that motivate the de-trending process that we do not want the (D)CNN to absorb \citep{STORM_ECHO}. The transfer functions used to generate the lightcurves are drawn from three populations: 
\begin{enumerate}
    \item A distribution of $\Psi$ that come from a selection of ``basis" shapes, including rings, gaussians, and exponential decay profiles in velocity-delay space. 
    \item A distribution of $\Psi$ drawn from an analytic $\Psi$ for a face-on accretion disk. While the accretion disk signal likely does not extend to the BLR, we include this as it is another nice representative ``picture'' of how gas can respond to changes in the continuum, and in the hope that the model may be able to learn to disentangle contributions from the accretion disk that may contaminate BLR measurements and vice versa. The assumptions and formulations of this accretion disk model are shown in Section \ref{sec:KDM} in the \hyperref[appendix]{Appendix}.
    \item A distribution of $\Psi$ drawn from simple kinematic disk, cloud, and combined models of the BLR similar to those described in \cite{Long_2023} and \cite{Long_2025}.
\end{enumerate}
The full parameter ranges used to generate each class of 1D and 2D transfer function are summarized in Tables~\ref{tab:1Dparams} and~\ref{tab:2Dparams} in the \hyperref[appendix]{Appendix}.
\begin{figure}
    \centering
    \includegraphics[width=0.48\textwidth,keepaspectratio]{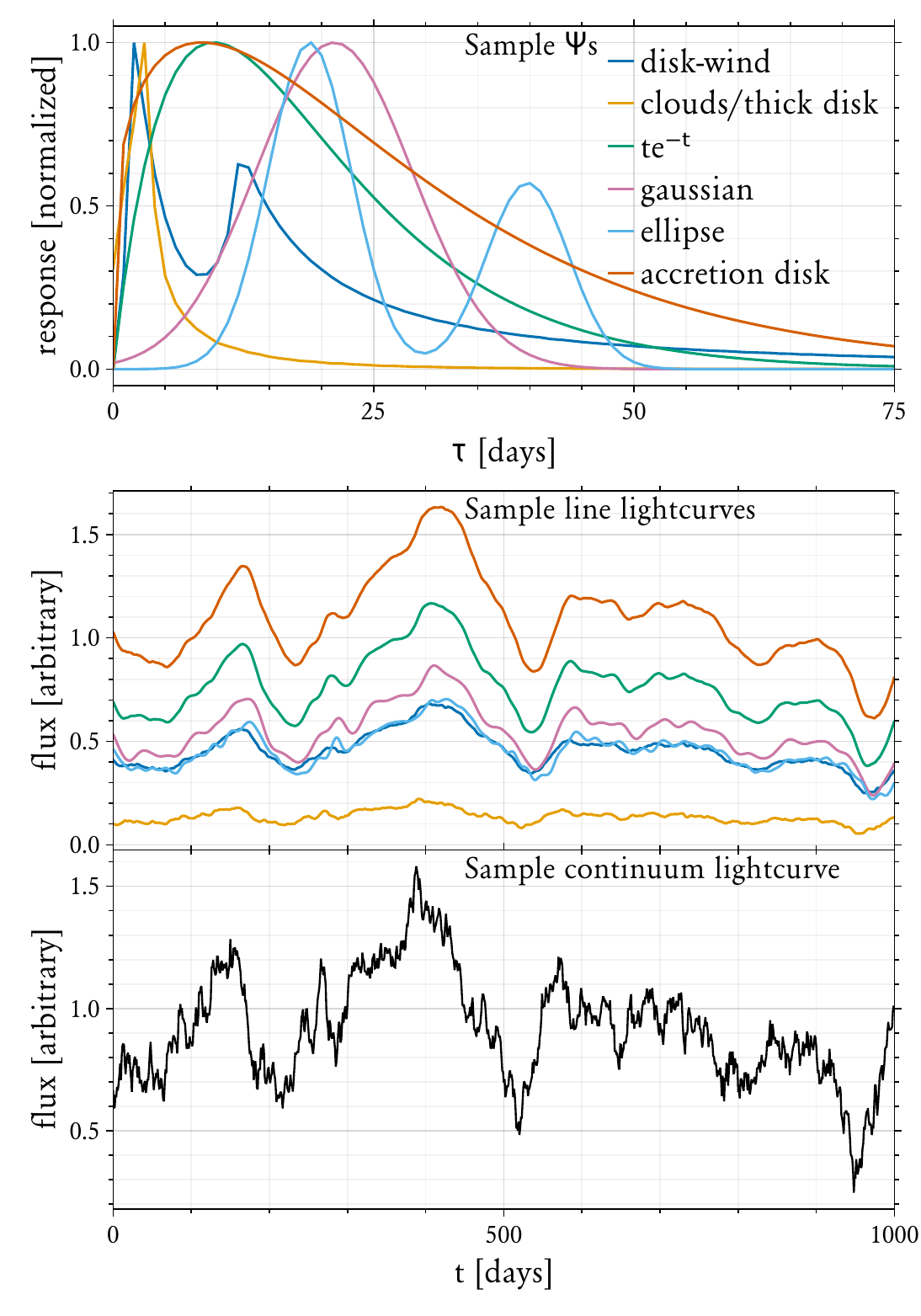}
    \caption{\textbf{Top}: Sample 1D transfer functions for each ``basis" function used in training the model. \textbf{Middle}: Sample lightcurves resulting from the convolution of the corresponding transfer function from the top panel with the continuum. \textbf{Bottom}: The DRW continuum lightcurve convolved with the top panel to generate the middle panel.}
    \label{fig:1Dtrain}
\end{figure}
In addition to generating samples independently, all three of these sources of $\Psi$ are also randomly added in combination to each other to generate additional samples, such that one lightcurve could be generated from say a basis shape alone or a basis shape in combination with a kinematic BLR model or a basis shape in combination with both a kinematic disk model and an analytic accretion disk model. The training set is generated such that the gaussian, ellipse, exponential, and accretion disk samples represent $1/16$ of the total training samples each, the BLR and disk-wind choices represent $1/8$ of the total training samples each, and examples that are random combinations of those existing subsets represent $1/2$ of the total training samples.\footnote{The full training sets are $\sim 50$ GB per scenario and thus cannot be cheaply hosted on the internet at this time, but all are available on request from the author.} 

All the resulting transfer functions are normalized, as we are asking the model to predict their shapes and not their intensities relative to each other. This is because we train on simple geometric, kinematic, and pictorial prescriptions for the BLR without detailed gas physics, thus our normalized model predictions can only place constraints on the geometry and kinematics of the BLR and not further constrain properties of the gas. In the future including training samples with more realistic radiative transfer effects would allow us to also recover the true amplitudes of the response, which is a key goal for future work but outside of the scope of this demonstration paper. One other important thing to note here is that the (D)CNN architecture requires that all inputs have the same size, thus when feeding it real data after training one must also match that expected shape. This means that either the training set should be generated to be the same expected size as the data or that one will have to interpolate the real data to be the same size as the training set lightcurves were. While we believe this training set represents good coverage of plausible BLR transfer function shapes, there are of course many models of the BLR in the literature which we have omitted for simplicity and ease of computation but could include in future iterations of this idea \citep[i.e.][to name just a few]{WATERS16,Mangham2019,ChajetHall2013,Flohic2012}.

We then train an ensemble of $\sim$ 100 (D)CNNs on our fake data, splitting it into 70\% training and 30\% testing data subsets. Each model is initialized with random Glorot normal \citep{glorot_2010} weights and is trained using the Adam optimizer with a learning rate of $\sim10^{-4}$, a value found experimentally to produce good results. We use a simple mean square error loss function to evaluate each model during training, which is evaluated with gradient descent optimization until each model arrives at some local minimum in the loss function. Each model is trained for a variable number of epochs, with the training automatically stopping when one of the following conditions is reached: 
\begin{enumerate}
    \item We compare the average training loss of the last five epochs with the average loss of the five before that. If there is no change or the average is increasing we stop training.
    \item We perform a similar test for the loss on the validation set and similarly stop training if the average validation loss across the five window average flattens out or increases, as this indicates the model may be overfitting to the training data. Note that although we evaluate the performance of the loss function on the validation data the model never sees this data in each gradient descent update, thus the training is blind to the validation set except in this autostopping routine. 
\end{enumerate}
During training we create checkpoints of our model state every time the validation loss reaches a previously unseen low, and reverts to the last saved checkpoint in case training is stopped before the maximum number of epochs is reached. The choice of using the previous five epochs in this way for the autostop criteria is arbitrary and was chosen experimentally as it provided a good balance of allowing enough chances for the model to ``escape" a local minima it may have stalled out in versus wasting computational resources when it truly has found the best solution for its initial conditions and optimization rate. In general we find that each model trains for $\sim 50$ epochs before this autostop criterion is usually triggered. Example training and validation loss curves are shown in Figures \ref{fig:2Dcombined}, \ref{fig:2Dtransfer}, and \ref{fig:2Dtest} that demonstrate this autostop criterion visually. We tested learning rates in the range $10^{-2}$--$10^{-4}$ and found through trial and error that $1.5\times10^{-4}$ produced the best results. A complete summary of the (D)CNN hyperparameters is given in Table~\ref{tab:hyperparams} in the \hyperref[appendix]{Appendix}.

After each model in our ensemble has been trained we generate a distribution of their final validation losses and remove any models that are significant poorly-performing outliers by removing any models that perform worse than a standard deviation away from the mean validation loss. This cut is again chosen arbitrarily but is done to remove models that did not learn very much about the problem due to their initial conditions and/or learning rates as well as to penalize models that overfit the data without generalizing what they learned to the broader scope of the problem. In general in imposing this cut we usually end up keeping $\sim 20-40$ of the original 100 trained models.

While computationally more expensive than training a single model, this ensemble approach allows for better characterizations of the uncertainty on each prediction in addition to allowing the models to explore the various minima that may exist in the loss function, as there is no guarantee that there is a global minimum or that each model will converge to the same minimum. The simplicity of the model architecture allows for each model to be trained with relatively few resources\textemdash the individual model sizes are $\lesssim$ 30 MB and many of the results shown here were generated on a laptop with an NVIDIA 5070 Ti GPU with just $\sim$ 6000 CUDA cores. The most computationally expensive models are the 2D cases shown here as the gradient sizes become larger, but even these are easy to train on a single node of a high-performance computing cluster with a single A100 GPU.  

\section{1D test cases}\label{sec:1D}
First, we demonstrate the recovery of the one-dimensional transfer function $\Psi(\tau) = \int\Psi(v,\tau)\rm{d}v$. In this case we follow the approach outlined in Section \,\ref{sec:models} but we include only information in the delay dimension as any velocity-dependent effects are integrated out. A sample transfer function and line lightcurve from each possible ``pure'' distribution included in our training set is shown in Figure \ref{fig:1Dtrain}\textemdash in total we generate $\sim 1.5\times 10^4$ fake samples drawn from these distributions and their combinations. The resulting dataset is split 70/30, such that the model sees roughly $10^4$ total samples in training and we can then test its predictions on the remaining $5\times 10^3$ samples it is blind to (the validation set). In this scenario we assume a continuum and integrated emission-line lightcurve have been measured at a fixed cadence for $\sim$ 1000 times the step size, which we scale in our plots to take a fixed cadence of one day with a total time baseline of 1000 days. Because of this choice we do not allow any transfer functions shown to the model in training to have any significant response past $\sim$ 1/10th of the total time baseline ($\sim$ 100 days). This ensures the model only picks up on trends that should be possible to measure given the constraints of the observations.  
\begin{figure}
    \centering
    \includegraphics[width=0.48\textwidth,keepaspectratio]{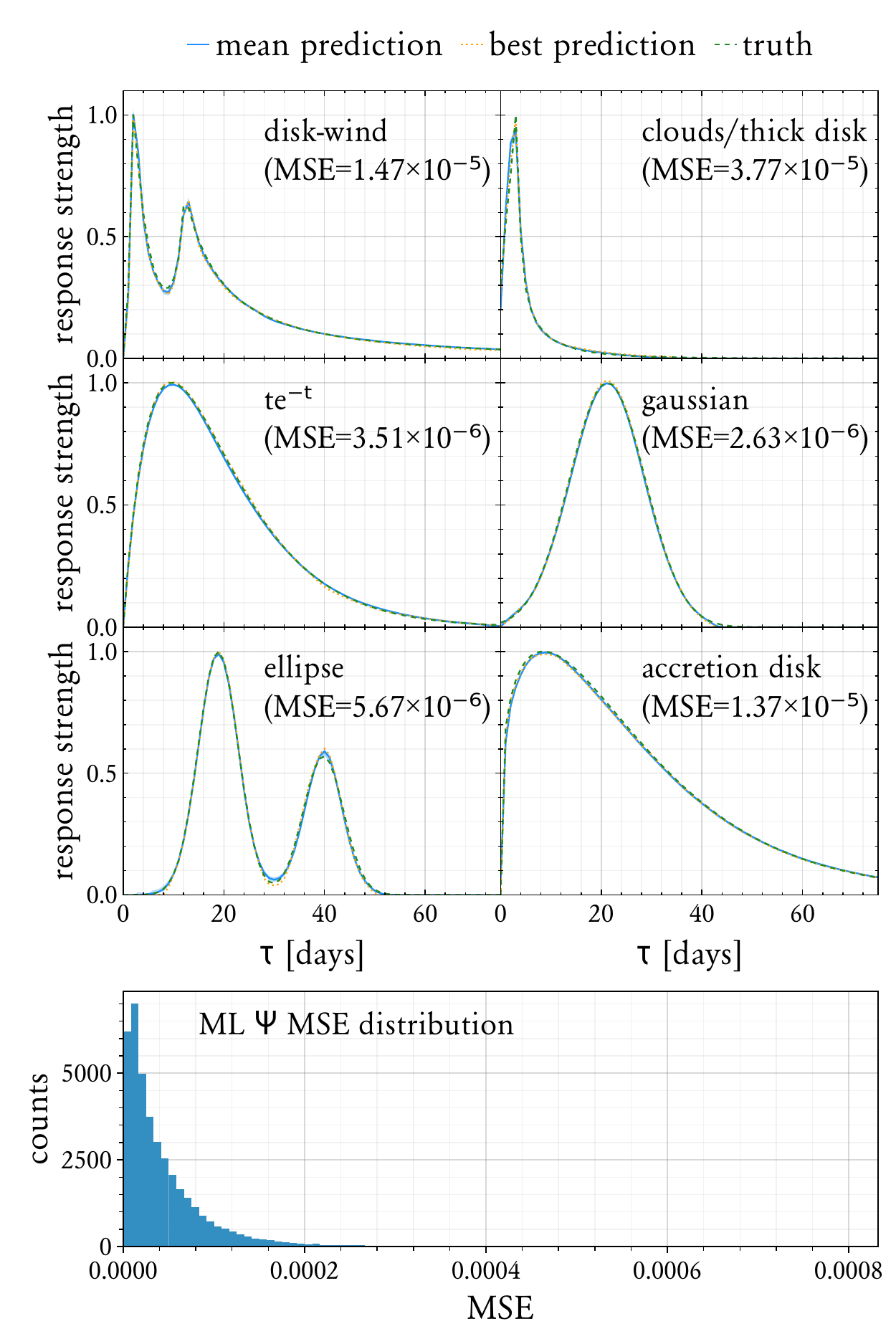}
    \caption{\textbf{Top grid}: Sample recoveries of each of the six ``basis'' transfer functions shown in Figure \ref{fig:1Dtrain}. \textbf{Bottom}: Distribution of errors for all of the model predictions in the validation set.}
    \label{fig:1DSampleResults}
\end{figure}

We start by training an ensemble of $\sim$100 1D (D)CNNs as described in Section \ref{sec:models}, and after training and imposing our cut to keep only the best performing models the final ensemble includes $\sim$ 20 trained (D)CNNs. The top panel of Figure \ref{fig:1DSampleResults} shows sample model recoveries for each distribution in the dataset, while the bottom panel shows the distribution of errors for all of the model predictions. Note that the best fit and mean fit lines are quite close to each other, indicating that all of the models in the ensemble are learning generalizable information and it is not just one model that happens to luckily find the optimal local minimum that matches the data. Also note that here just a single perfect lightcurve is fed to the models, thus the error bars are quite small and represent only the errors from the spread of the mean prediction of each member of the ensemble for this perfect input lightcurve.

\subsection{Missing/erroneous data}\label{sec:worseData}

\begin{figure*}
    \centering
    \includegraphics[width=\textwidth,keepaspectratio]
    {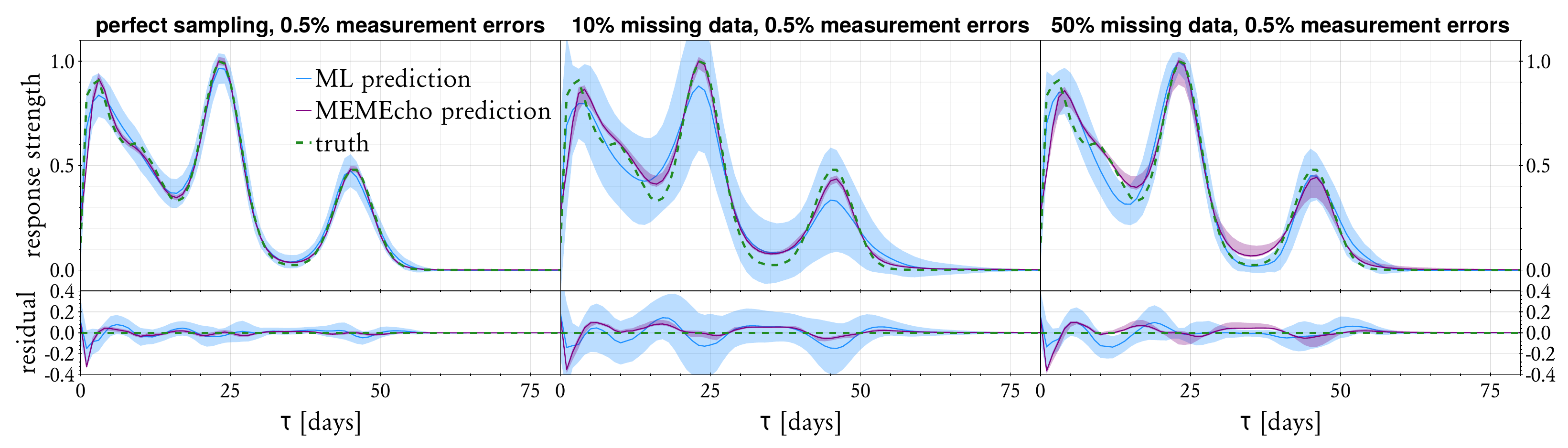}
    \caption{Comparing how the novel (D)CNN and existing analytic \texttt{MEMEcho} techniques perform when the training data quality are degraded \textbf{Left}: Recoveries of a more complicated 1D transfer function that is the result of the combination of several of the ``basis'' functions shown in Figure \ref{fig:1Dtrain} and Figure \ref{fig:1DSampleResults}. The top panel shows the (D)CNN recovery with error envelope in blue, the \texttt{MEMEcho} recovery with error envelope in purple, and the ground truth as the dashed green line. The bottom panel shows the residuals with the same colors. \textbf{Middle}: Sample recoveries of the same transfer function with the same colors as the left panel, but this time from data with random observational gaps equaling 10\% of the total number of data points. \textbf{Right}: Sample recoveries of the same transfer function with the same colors as the left panel, but this time from data with random observational gaps equaling 50\% of the total number of data points.}
    \label{fig:1DMissingResults}
\end{figure*}

While the results above already include noisy observations, they are perfectly sampled with a set cadence in time space. An important question is then how the model predictions might degrade given gaps in observations, and to test this question we train the models as before but this time remove 10\% of all the time samples. As expected, this does degrade the quality of the predictions somewhat, particularly in resolving sharp features, but overall the models still learn an impressive amount of generalizable information and recover the transfer functions well. For example, when the input lightcurves are just noisy but perfectly sampled as shown in Section \ref{sec:1D} the average MSE is $\sim 5\times 10^{-5}$, and when sampling gaps are introduced this average error increases to $\sim 1.6\times 10^{-4}$, a factor of $\sim 3$ increase, a ratio that is similar in both 1D and 2D inversion tests. In Figure \ref{fig:1DMissingResults} we showcase how degrading the training data quality affects the recovered 1D transfer functions. As expected the 
recoveries are noisier and less precise, but it clearly can still recover overall trends, and importantly in both the perfect and missing data cases the error bar always encapsulates the true response function. Interestingly, the (D)CNN predictions remain consistent even when removing as much as 50\% of the input data, as shown in the far right panel of Figure \ref{fig:1DMissingResults}. The right panel's predictions come from the same set of models used in creating the middle panel, trained on just 10\% missing data yet predicting reasonably well even when 50\% of the data have been removed. This behavior does not extend to the models used in creating the left most panel, which were trained on perfect data\textemdash feeding them any missing data causes their prediction quality to degrade significantly, thus it is critical when training (D)CNNs to include data dropouts in the training samples.

This inversion problem has existed long before machine-learning methods were a tractable possible solution, and the \texttt{MEMEcho} \citep{MEMEcho1,MEMEcho2} code has existed as the most widely-used analytic approach to tackling this inversion problem. Thus in Figure \ref{fig:1DMissingResults} we also show for comparison the \texttt{MEMEcho} inversions for the same data. As Figure \ref{fig:1DMissingResults} shows, both the \texttt{MEMEcho} and (D)CNN inversion are nearly perfect when given nearly perfect data as expected, but in degrading the data quality they both begin to diverge slightly from the ground truth. Note that in generating the far right panel's \texttt{MEMEcho} predictions it is shown an emission line lightcurve with 50\% missing data points but a continuum lightcurve with just 10\% missing data points, as the equivalent (D)CNN prediction comes from an ensemble of models trained on 10\% missing data.

The error bars on the (D)CNN model predictions are generated by having the ensemble generate $10^4$ predictions from as many distinct, Monte Carlo sampled lightcurves, which we then use to calculate a mean as well as upper and lower standard deviations on that mean to show in the ribbon. This confidence interval is calculated as the sum in quadrature of the spread of ensemble model means and the spread of each individual model when fed a large enough number of lightcurves. In general we find that the contribution to the error budget is dominated by this Monte Carlo process, and without this the error bars shrink by a factor of $\sim 5-10$. Note that other common error estimation procedures such as bootstrapping are difficult to get to work with the (D)CNN method, as the (D)CNN method requires consistently sized inputs and in our current formulation does not take in any information about the uncertainty of each point in the lightcurve, thus the only way to account for this is to feed it many lightcurves through Monte Carlo sampling. To generate the error bars on the \texttt{MEMEcho} fits we ran \texttt{MEMEcho} $10^4$ times on fake data generated in the same way as it was for the ML model predictions, with the ribbon corresponding again upper and lower standard deviations on the mean \texttt{MEMEcho} fit. We use a target $\chi^2$ level of $1$, keep the time sampling restricted to $\sim 1$ day, and iterate each fit for a maximum of $10000$ iterations, but otherwise use default \texttt{MEMEcho} values, thus expert tinkering may be able to further improve the \texttt{MEMEcho} recoveries shown here.

This is one advantage of the (D)CNN methodology\textemdash when training an ensemble of models it is easy and cheap to generate uncertainties directly, whereas while an individual run of \texttt{MEMEcho} is relatively fast generating uncertainties is much slower as the algorithm must be run many times over. It is encouraging that the (D)CNN method error bars look realistic in that the models are more unsure when the data quality is worse, and the places it is most unsure are where there are sharp features in the data. Importantly, in all three cases the ground truth is nearly always captured by the (D)CNN error bars. 

We also test our models for systematic biases across the entire test set by comparing how accurate the mean delays ($\tau$ centroid, first moment) and delay widths (second moment) from the ground truth vs. the (D)CNN ensemble prediction. In doing so we see that our 1D models almost never overpredict the mean delay $\bar{\tau}$ and are instead systematically biased slightly towards underpredicting $\bar{\tau}$. As expected, this is a more significant problem at larger $\bar{\tau}$, and is thus likely a limitation of the training set\textemdash which have lightcurves of fixed duration\textemdash and not of the method itself. We find that the models also systematically underestimate the widths $\sigma_\tau$ of the transfer functions, and that this is correlated with the underprediction of $\bar{\tau}$. A full discussion on the systematic biases as well as plots showcasing the regimes where the models are likely to have biases is available in Section~\ref{sec:bias} in the \hyperref[appendix]{Appendix} (Figures~\ref{fig:bias1Dscatter} and \ref{fig:bias1Dhist}).

\section{2D test cases}\label{sec:2D}

While the 1D results are encouraging, modern instrumentation and data analysis techniques have enabled two dimensional reverberation mapping campaigns, where the lightcurves are measured as a function of velocity across the emission-line. Such 2D datasets provide new opportunities to constrain the fundamental nature of the broad-line region and reveal important information that is missed in recovering just the integrated 1D response \citep{Long_2023}. Obtaining the most accurate full 2D echo maps is thus critical to unraveling the true nature of the broad-line region.

\begin{figure}
    \centering
    \includegraphics[width=0.48\textwidth,keepaspectratio]{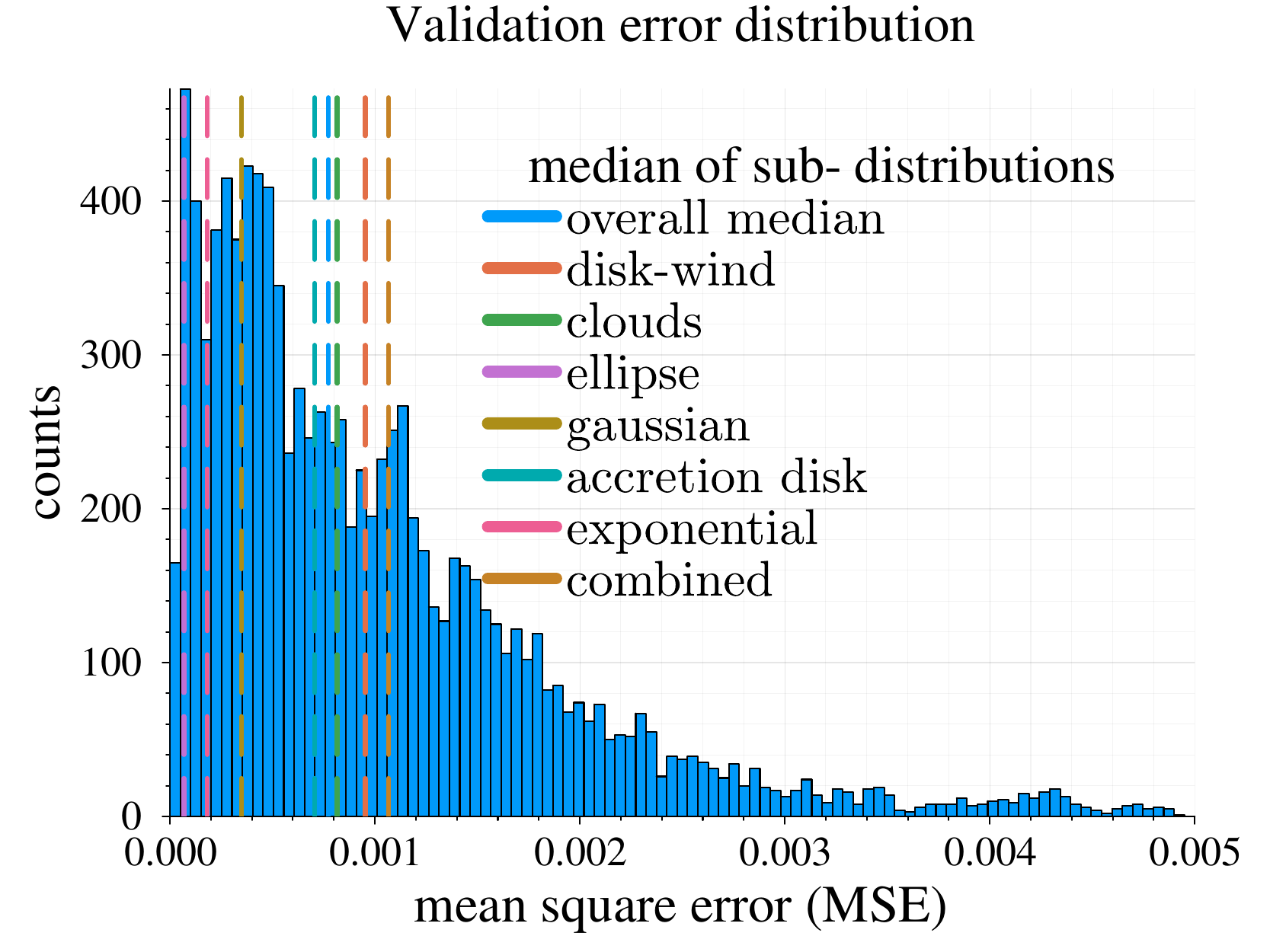}
    \caption{The distribution of errors for the sample ensemble of 2D models. Full examples of each model transfer function and lightcurve inputs are in Figure \ref{fig:2DSamples} in the \hyperref[appendix]{Appendix}, and a sample inversion for a complicated hybrid BLR is shown in Figure \ref{fig:2Dcombined}. Vertical lines showcase the median error of each sub-population for each of the basis functions.}
    \label{fig:2Derrordist}
\end{figure}

Here we consider how well our (D)CNN technique works for the full 2D case as well, performing similar tests as shown in Section \ref{sec:1D}. We again train an ensemble of  models following the approach outlined in Section \ref{sec:models}, but this time use full 2D transfer functions to generate lightcurves at 25 distinct velocity bins, examples of which are shown in full in Figure \ref{fig:2DSamples}  in the \hyperref[appendix]{Appendix}. The distribution of errors for the full 2D cases are shown in Figure \ref{fig:2Derrordist}. Following the lessons learned from our 1D tests, we train all of our 2D models on 10\% missing data. 

\begin{figure*}
    \centering
    \textbf{Visualizing a more ``complicated" transfer function inversion}
    \includegraphics[width=\textwidth,keepaspectratio]{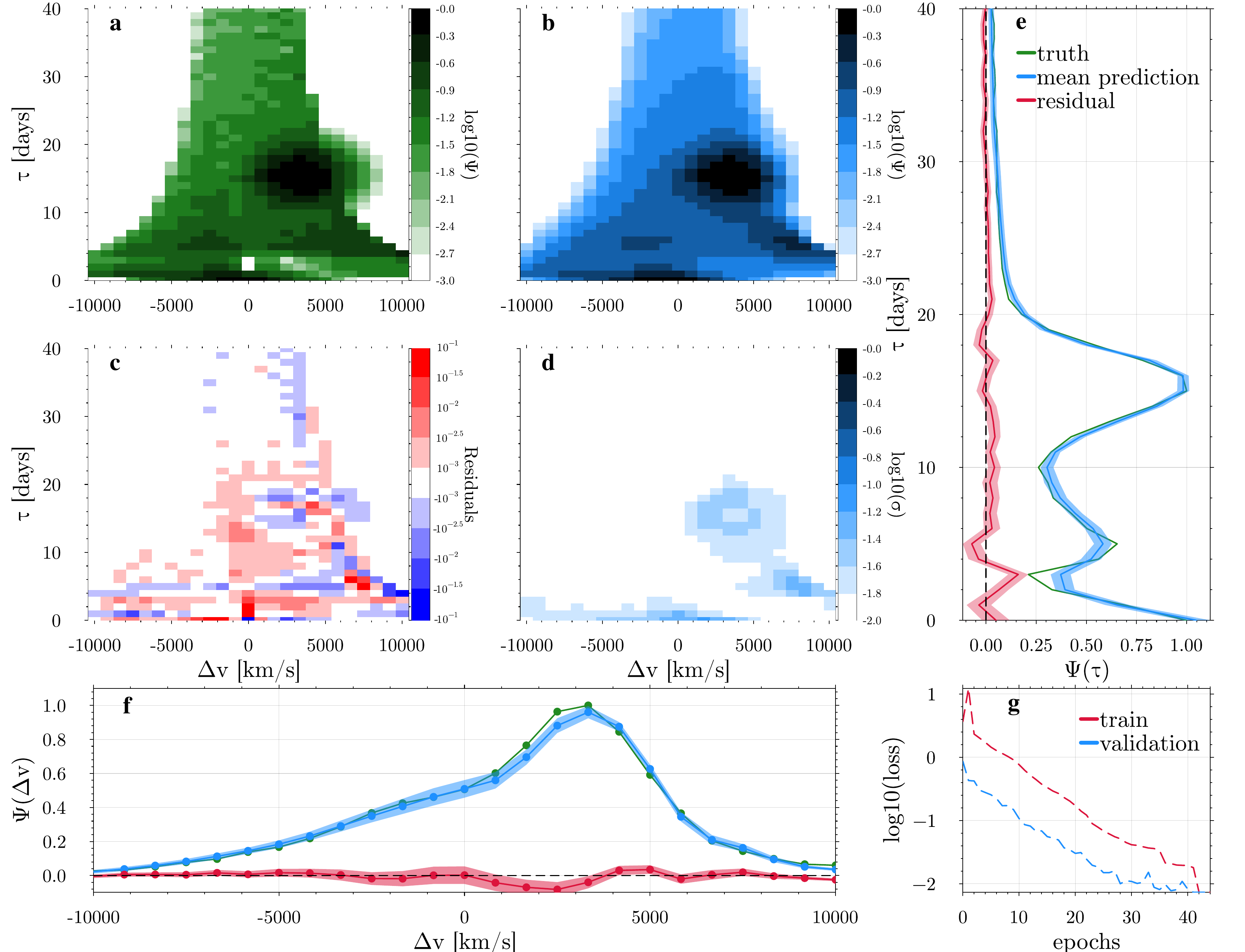}
    \caption{\textbf{a}: Sample full 2D transfer function of a complicated combined synthetic BLR with both a ``blob'' component superimposed on a more traditional virial ``disk-wind'' component. \textbf{b}: (D)CNN model recovery of the full 2D transfer function. \textbf{c}: Residuals (prediction in top middle panel - ground truth shown in top left panel). Regions where the model overpredicts are shown in red, underpredicted regions are shown in blue, and areas in white are where the model prediction is within $\pm0.1\%$ of the maximum ground truth value. \textbf{d}: The 2D uncertainty in the ensemble of predictions for this lightcurve. \textbf{e}: The 1D transfer function (2D maps integrated over velocity, see Figure \ref{fig:1DSampleResults}) showing the mean model prediction in blue (with ribbon corresponding to uncertainty), the ground truth in green, and the residuals in red. \textbf{f}: The 1D line profile (2D maps integrated over delay) again showing the mean model prediction in blue (with ribbon corresponding to the uncertainty), the ground truth in green, and the residuals in red. \textbf{g}: Mean loss curves for the ensemble of models, with the validation shown in blue and the training set shown in red. Note that the models stop learning after $\sim$ 30-40 epochs.}
    \label{fig:2Dcombined}
\end{figure*}

Unfortunately in training the gradients become much larger and more difficult to compute in the 2D case, so we restrict the model depths to just one split/skip layer as shown in Figure \ref{fig:model}, which reduces the model sizes to just $\sim$ 10 MB. To start, we again use data with measurement errors but full time-sampling. Given the complexity of the problem and to save page space, instead of showing sample recoveries for all basis functions and a complicated case we instead show just one showcasing a more complicated combined BLR in Figure \ref{fig:2Dcombined} as the 2D analogue for the ``more complicated'' 1D transfer function recoveries shown in Figure \ref{fig:1DMissingResults}. This more complicated transfer function is a combination of a largely ``disk-wind'' virial BLR with a gaussian ``blob'' off-center and at longer delays. This creates a complicated structure in velocity-delay space that the model recovers qualitatively quite well. 

\begin{figure*}
    \centering
    \textbf{2D transfer learning example}
    \includegraphics[width=\textwidth,keepaspectratio]{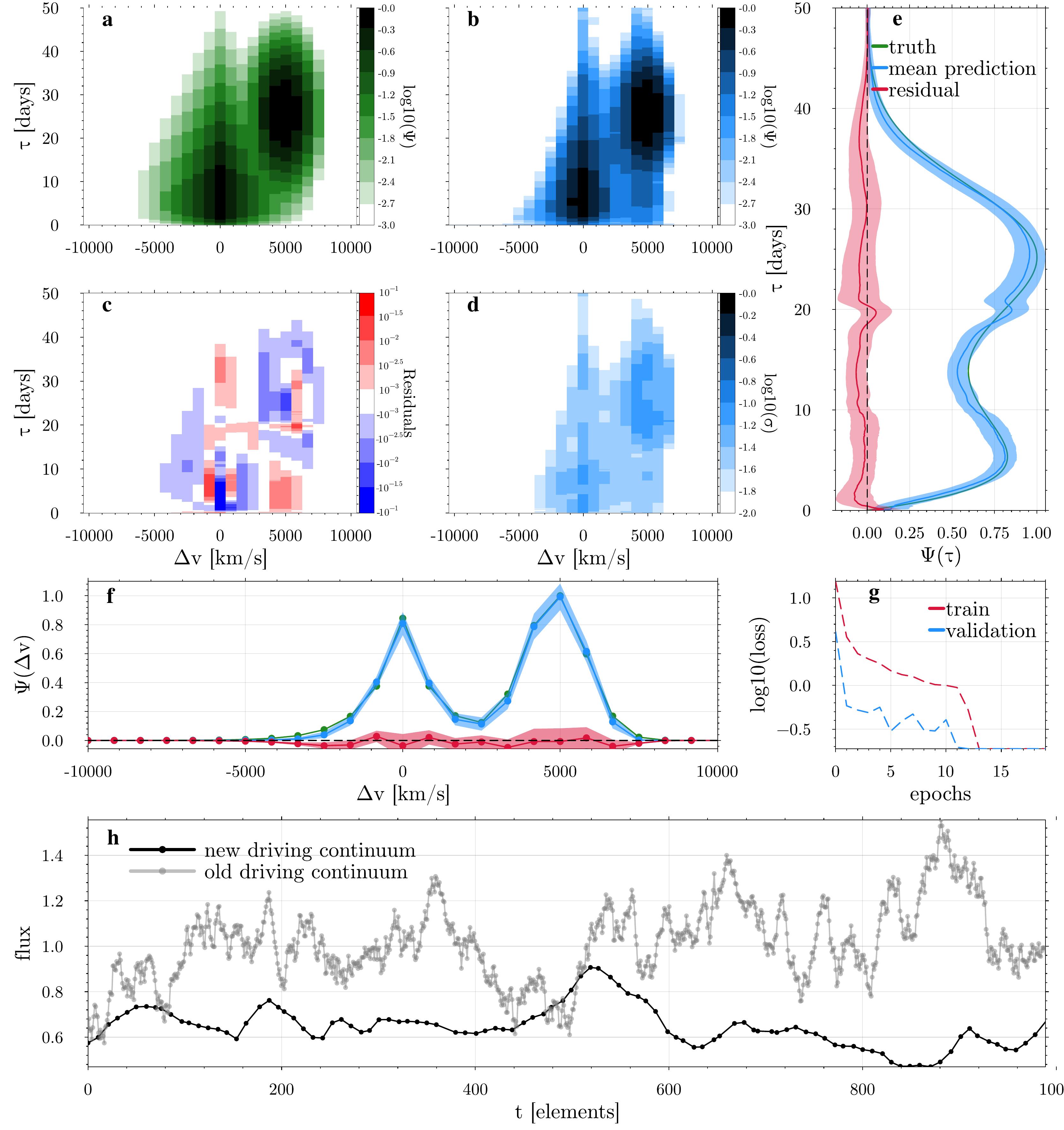}
    \caption{A 2D inversion of another complicated combined transfer function, this time the result of a two ``shape'' BLRs (an exponential decay component + a gaussian blob). Panels \textbf{a-g} are the same as described in the caption of Figure\,\ref{fig:2Dcombined}. Panel \textbf{h} shows the driving lightcurves the (D)CNN models transfer learned between. The old driving continuum shown in lighter grey is what was used to train the models shown in Figure \ref{fig:2Dcombined} and the new driving continuum shown in black is what the models adapted to in order to make the predictions shown here. Note the x-axis of this panel is in ``elements'' as this new continuum lightcurve has a reduced physical timescale ($\sim 100$ days instead of $\sim 1000$ days) and was thus stretched by interpolation to match the old input lightcurve size. The original data points (non-interpolated) for this lightcurve are shown with markers.}
    \label{fig:2Dtransfer}
\end{figure*}

The inversion shown in Figure \ref{fig:2Dcombined} has a mean square error of $\sim 10^{-3}$, making it quite close to the median error in the distribution shown in Figure \ref{fig:2Derrordist} and thus representative of the quality of possible predictions for this (D)CNN approach. Qualitatively the model does a good job in identifying where the distribution of gas lies around the black hole when comparing panels a and b showing the ground truth and model prediction, respectively, and appears to mostly blur/smooth things a bit when compared to the ground truth.  Where the model is led astray is clear in the middle left and middle center panels\textemdash in the residuals it is clear that the mean prediction gets the width of the virial envelope wrong by a $\sim$ pixel, and this sharp edge in the ground truth results in higher errors near these edges. The $\sigma$ panel (d) shows that the model is in general most uncertain where it is most wrong, which is an encouraging diagnostic as this means if we have no access to the ground truth we would still have helpful information about which portions of the inversion were the least reliable.

The 1D projections (panels e and f) show that the inversion quality has not decreased significantly from the original 1D results presented in Section \ref{sec:1D}, despite having a much more complicated architecture through the introduction of the second dimension in the data quality. Note that here the uncertainty ribbons are underestimates, as this is just the uncertainty from the ensemble of $\sim 30$ models applied to the same input lightcurve. This isolates the uncertainty in the (D)CNN models themselves, but in practice one would also feed many realizations of possible lightcurve inputs given their measurement errors which would increase these uncertainties. 

As Figure \ref{fig:2Derrordist} shows, in general the ``shape" basis functions are easiest for the model to learn and the ``physical" models the hardest, particularly in the case of the ``disk" type inversions. This is likely due to finer structural changes in these classes of models leading to scenarios like that described above in the interpretation of Figure \ref{fig:2Dcombined}, but overall the distribution between the classes is relatively tight and each is predicted well. 

The bottom right panel (g) of Figure \ref{fig:2Dcombined} verifies that our models do not overfit the data and our training procedure as outlined in Section \ref{sec:models} is working as intended. In particular it is encouraging to see that the morphology of the validation and training curves is nearly identical, which indicates that the models are learning strongly generalizable information about the inversion problem. Note that the relative offset between each curve is a byproduct of the fact that the training set contains more data (70\% of the total) than the validation set (30\% of the total). 

This is just one example inversion, thus we also compute statistics across the entire test set to test if there are any systematic biases. In doing so we find that the
models almost never overpredict delays, and instead are systematically biased towards underpredicting delays slightly, especially at longer delays as one would expect.
In velocity space we find no systematic biases. A deeper discussion on these biases and plots showcasing their distributions are available in Section~\ref{sec:bias} in the \hyperref[appendix]{Appendix} (Figures~\ref{fig:bias2Dhist} and \ref{fig:bias2Dscatter}).

\section{Transfer learning}\label{sec:transfer}

All the results thus far have essentially been an exercise in model fitting, where the (D)CNN has adapted itself to fit one particular continuum lightcurve. While this is valuable alone, ideally our model should be able to adapt to a variety of continuum lightcurves. While any particular model at the end of training will be optimized to the continuum lightcurve used in generating its training samples, it turns out that it is quite easy to get the model to adapt to a different continuum lightcurve through \textit{``transfer learning''}. 

To do this we generate a new training dataset in the same way as before but with a new continuum lightcurve. This new continuum lightcurve is markedly different than the DRW lightcurves used previously, as it is obtained from synthetic quasar spectra first published in \cite{Mangham2019}. The fake observational campaign of this quasar is sampled at a $10\times$ lower rate than our previous assumptions, thus we interpolate the obtained lightcurve to ensure the dimensionality of our inputs matches our previously trained model sizes. We then train the ensemble of models again but this time instead of initializing their weights randomly we start with their previously learned parameters from the old continuum.

In this way the models are already much closer to the solution than they are when initialized randomly, and in our tests they take only a few epochs of training with many fewer training samples to transfer what they learned from the old continuum to the new one, thus updating the model for a new continuum is a relatively quick and easy task when compared with the initial training requirement. In this way it is better to think of this method not as a general catch-all approach for every possible source and transfer function, but instead as a first test of a novel fitting method for each particular source that can interpolate in a highly non-linear way between the supplied transfer functions it has seen in training to generate a flexible best fit. 

The results of this process are demonstrated in Figure \ref{fig:2Dtransfer}, where we took the parameters of the ensemble used to generate the inversions in Figure \ref{fig:2Dcombined} and fine-tuned them on a new training data-set to generate the inversion shown for another complicated transfer function. This time we show a recovery of a hybrid ``shape'' BLR, where there is a strong exponential component and a gaussian blob again offset from line-center. Despite the complicated nature of this picture and the dramatically different continuum, the (D)CNN ensemble still recovers well both the full 2D distribution and the 1D projections.

Note that the loss drops rapidly and the training could have been stopped after just a few epochs to produce a similar quality inversion, showcasing how easy it is to transfer learn to new lightcurves. The fake data used in training for this problem are generated using a much noisier continuum, as it was extracted by hand from the simulated set of fake observational spectra first published in \cite{Mangham2019} to simulate how one would apply this approach to real data where the intrinsic continuum is not so precisely known. Additionally the data quality are degraded further as described in Section \ref{sec:worseData} by including gaps at 10\% level similar to the 1D test shown in Figure \ref{fig:1DMissingResults}. Here it is also harder to discriminate between the best and worst performing models, and our $1\sigma$ cut results in the largest ensemble in the paper\textemdash retaining $\sim 40$ of the original $100$ models.

This transfer learning process could  be repeated by any end-user of our (D)CNN models to fine-tune to their needs\textemdash for example if one wanted to include a new basis function in the model training set one would simply generate the new samples for this basis function and then fine-tune the model on those new samples starting from its previously trained position as published here.

\begin{figure*}[!ht]
    \centering
    \textbf{Testing the (D)CNN method on a more realistic BLR inversion problem}
    \includegraphics[width=\textwidth,keepaspectratio]{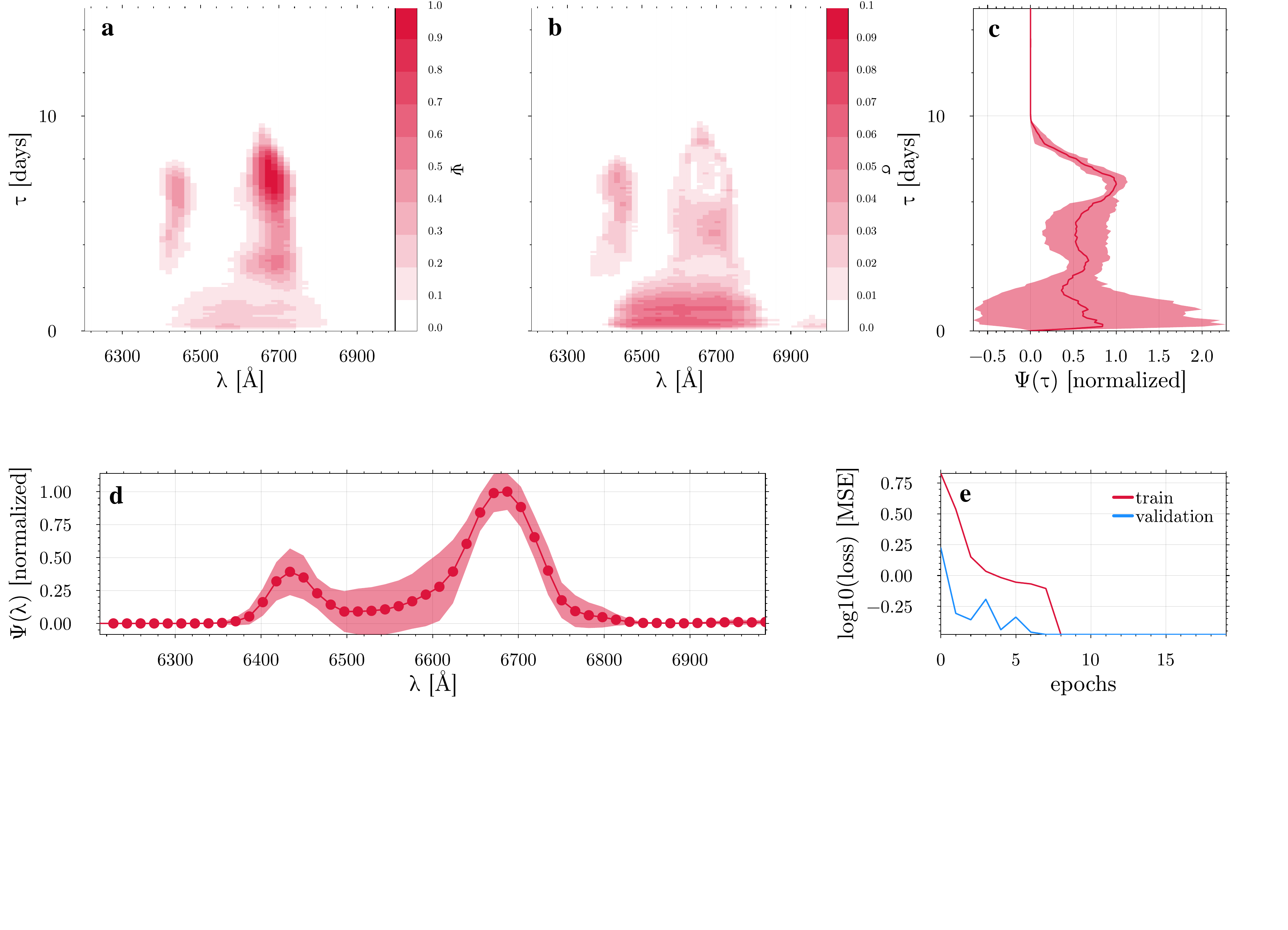}
    \vspace{-35mm}
    \caption{(D)CNN inversion of lightcurves extracted from a simulated quasar BLR dataset first presented in \cite{Mangham2019}. \textbf{a:} mean predicted $\Psi(\lambda,t)$. \textbf{b:} uncertainty on mean prediction. \textbf{c:} 1D transfer function $\Psi(\tau)$ with error bar corresponding to $\pm1\sigma$ uncertainty. \textbf{d:} 1D $\Psi(\lambda)$ with error bar corresponding to $\pm1\sigma$ uncertainty. \textbf{e:} mean loss curves for the ensemble of models.}
    \label{fig:2Dtest}
\end{figure*}

The goal of this transfer learning is to demonstrate how one might apply these models to real data, and to do this we perform a final test of our transfer-learned 2D (D)CNN models to see how well they can recover response functions from more realistic data. To accomplish this we follow the ``blind" test first performed with CARAMEL and \texttt{MEMEcho} in \cite{Mangham2019}. 

While the experiment has since been unblinded when \cite{Mangham2019} was published, we do not include anything like the full MHD + radiative transfer calculations that go into the more realistic BLR models shown in \cite{Mangham2019} in our training set, and we use only the series of synthetic spectra like one would from a telescope thus we preserve the spirit of the blind test well. We extract continuum and line lightcurves from these synthetic spectra in the same way one would for use with \texttt{MEMEcho}, and use the resulting continuum lightcurve to generate the transfer learning dataset. We transfer learn 25 independent models on this dataset and keep the top 5 of them after imposing our $1 \sigma$ cut. We add noise to the line lightcurves extracted from the \cite{Mangham2019} spectra at 1\% the extent of the variations to generate a set of $10^4$ noisy lightcurves to use as inputs for every model in our ensemble, yielding a total of $5\times10^4$ predictions and allowing us to calculate confidence intervals.

\begin{figure}
  \centering
  \vspace{3mm}
  {\includegraphics[width=0.48\textwidth,keepaspectratio]{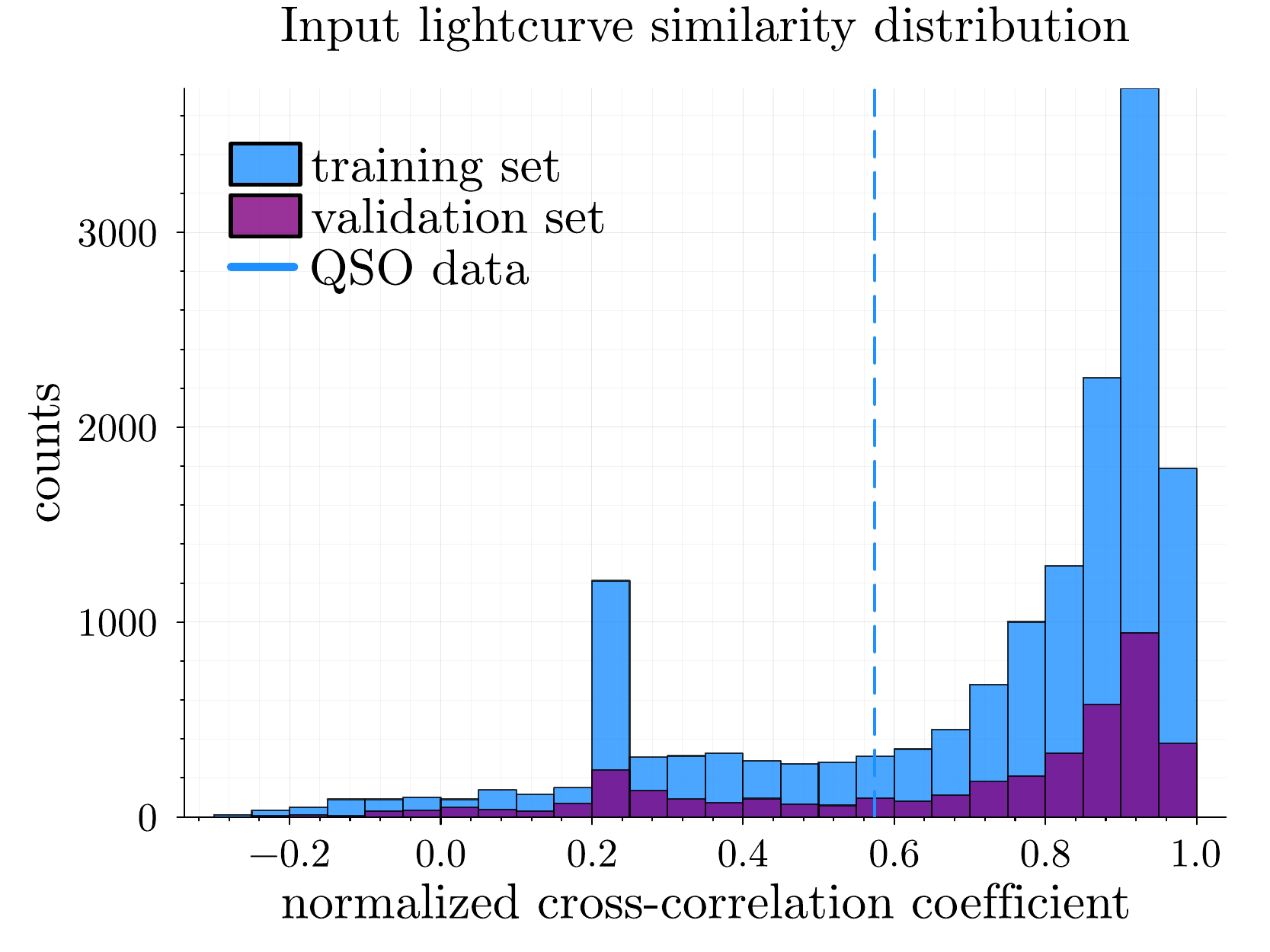} 
  \label{fig:similarity}}
  \vspace{-5mm}
    \caption{Similarity distributions of the lightcurves used in training/validating the model, with the vertical line indicating how the set of extracted lightcurves for the quasar BLR in \cite{Mangham2019} compares to the distribution used in training.}
  \label{fig:similarityQSO}
  \vspace{-3mm}
\end{figure}
The results of this ``blind test" are shown in Figure\,\ref{fig:2Dtest}, which shows how well the (D)CNN method recovers the shape and distribution of response for this test case. The true response function is shown in Figure 6 of \cite{Mangham2019}\textemdash the (D)CNN correctly identifies the velocity and time locations of peak response and the red-blue asymmetry in the response, but predicts extra power at small delays, underpredicts at longer delays, and does not resolve the features as sharply as they are present in the ground truth. The (D)CNN prediction is thus consistent with the existing \texttt{MEMEcho} recovery shown in Figure 17 of \cite{Mangham2019}, with the largest difference being that the \texttt{MEMEcho} recovery does not show the extra prompt response in the (D)CNN recovery. This extra prompt response near 0 days is physically interesting as this is likely the result of not having entirely subtracted out the continuum signal properly from the line lightcurves, as the training set we use has no continuum contamination. To test this theory we performed an inversion on the lightcurves that were not continuum subtracted\textemdash which are thus dominated by the continuum\textemdash and found that the models only predict a signal similar to this prompt response signal that remains when making a prediction on the de-trended lightcurves. This highlights how important properly de-trending the lightcurves is, as failing to do so results in the continuum variations dominating the inferred response. In the future we could adjust the training set to be more realistic, including contamination from the continuum, and believe this would likely help resolve this ``problem" and make our models less sensitive to this bias. It also showcases that this method has promise outside of applications directly to the BLR, and could likely be applied to RM in the accretion disk and torus \citep{RM_Review_21}.

Figure \ref{fig:similarityQSO} also highlights another potential way to test how confident one should be in a (D)CNN prediction. As shown in the bottom panel of Figure \ref{fig:similarityQSO}, the set of input lightcurves used in generating the (D)CNN prediction are significantly outside the peak of the distribution for what the model was shown in training. This indicates another potential area for improvement as well as a providing a qualitative way to benchmark trust in predictions\textemdash if the models had seen many similar sets of lightcurves in training we may be more confident in its output prediction and vice versa, and additionally we may improve the performance of our models by finding a way to generate samples that are statistically more similar to the data we wish to predict, but this outside the scope of this paper and we leave it for future work.




\begin{figure}
    \centering
    \includegraphics[width=0.48\textwidth,keepaspectratio]{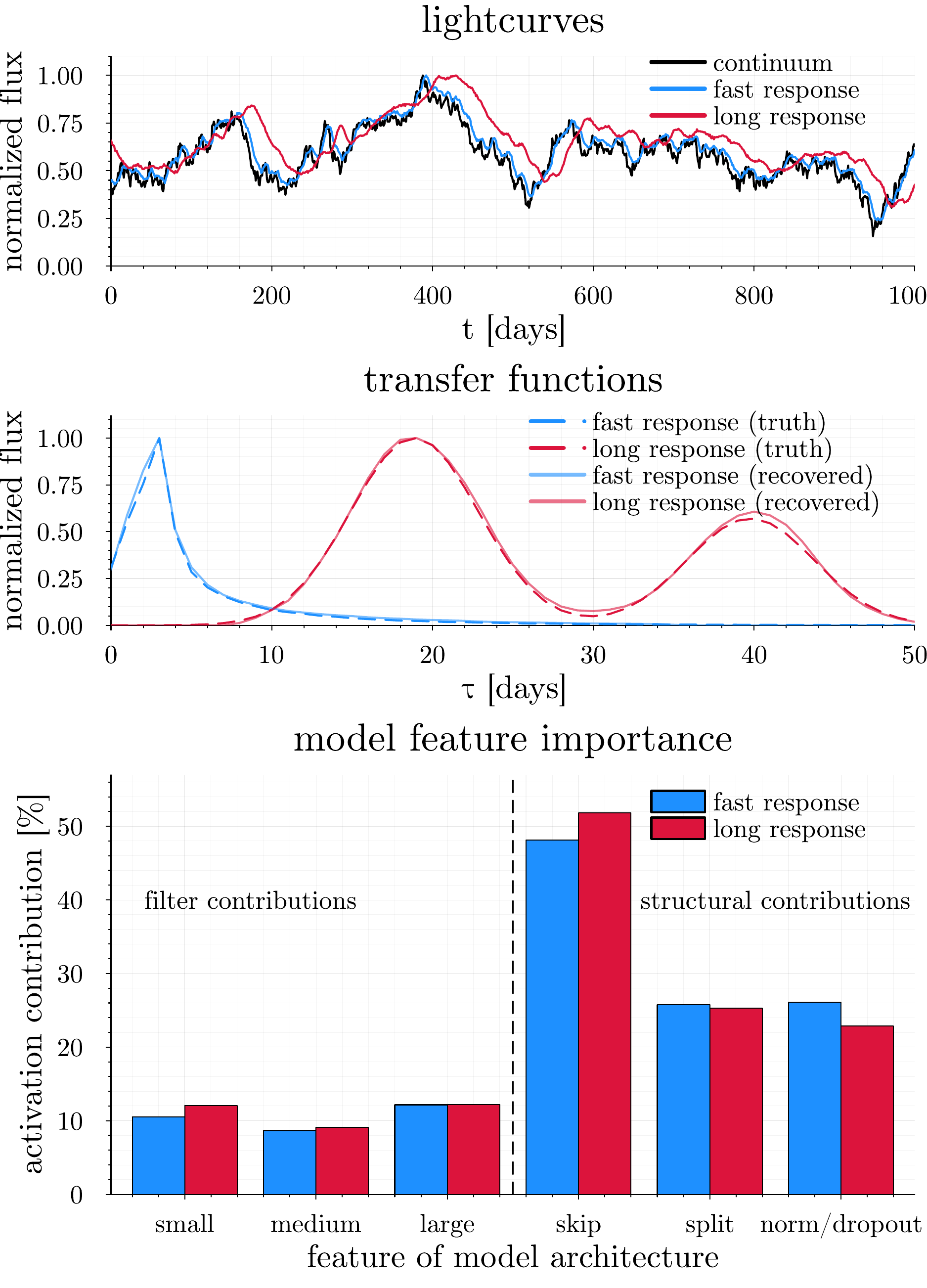}
    \caption{Here we showcase the percentage of activations (how strongly the model responds to an input and thus how much each feature contributes to the final prediction) for two very different input lightcurves.
    \textbf{Top:} The input lightcurves are shown in blue and red, while the continuum is shown for reference in black.
    \textbf{Middle:} The transfer functions used to generate the lightcurves in the top panel by convolving with the continuum. The blue is the ``cloud" transfer function and the red is the elliptical transfer function as shown in Figure \ref{fig:1DSampleResults}. The narrow peak at short delay times in the blue produces the fast variations seen in the lightcurve, while broader and later extent of the red transfer function results in the longer variations in its corresponding lightcurve.
    \textbf{Bottom:} The activation percentages of various components of our model architecture. The left three bars (small, medium, and large) correspond to the sum of all activations in each of the corresponding time filters. The blue ``skip" bar showcases how important the overall skip connection design is for the model, while the purple ``split" bar showcases the importance of the branching path components. Finally, the black ``norm/dropout" bar demonstrates the importance of the normalization and dropout layers in making the prediction. Note that thus the right three bars add up to 100\% but the left three bars do not as they are subcomponents of the ``skip" and ``split" paths.}
    \label{fig:interpretability}
\end{figure}


\section{Interpretability}\label{sec:interpretability}

While the results above are encouraging demonstrations that the (D)CNN method can recover transfer functions from RM observations, one drawback of many machine learning methods is their interpretability. In designing the architecture of our CNN (see Figure \ref{fig:model}) we attempted to let relevant physical scales inform our choices of filters and layers, in the hope that this might make the resulting predictions more interpretable. For example, one might expect that a lightcurve that varies at very short timescales would see strongest activations in the layers of the network with small timescale features, and conversely an input lightcurve that varied slowly would see result in most of the model activations occurring in the layers that corresponded to the longest timescales. We test this idea in Figure \ref{fig:interpretability}, showing that this is distinctly \textit{not} how the model operates. 
There appear to be no discernable changes between the activation of different timescale filters even when the inputs clearly show such different time-varying behavior, making interpreting how the (D)CNN works indeed more of a ``black-box" than one might like. Interestingly the only trend between the two input lightcurves is the change in importance of the ``skip" and ``norm/dropout" layers, implying that the normalization and dropout layers appear to be more important for faster varying inputs. 

\section{Discussion and Future Work}\label{sec:discussion}
In this paper we have shown that a (D)CNN can successfully deconvolve reverberation mapping data products, both in 1D and in recovering the full 2D velocity-delay maps. The method works even in the face of significantly erroneous/missing data, and can naturally generate sensible confidence intervals on this prediction by training an ensemble of models. The predictions generated with this new approach also encouragingly agree with the results of analytic inversion methods such as \texttt{MEMEcho}. While we have designed this new method with the BLR in mind, in principle this technique could be applied to any reverberation deconvolution problem, including in the accretion disk and torus. Unfortunately even though the network's architecture was designed with physics in mind, simple tests to demonstrate the model behaves in the way we might intuitively expect show that its behavior is difficult to interpret.

Given the analysis in Section \ref{sec:interpretability} it may make sense for future iterations of (D)CNN model architectures to forgo the physically motivated ``split" layers, and instead use only the ``skip" connection architecture. This would likely save training time and make the models even more portable. Additionally, in training future models one could explore different loss metrics\textemdash while here we have shown that the standard mean square error metric works well, custom loss functions incorporating further knowledge of the specific problem at hand may allow for further learning at the cost of computational complexity. For example one could imagine using a version of an image similarity metric such as the normalized cross-correlation coefficient\textemdash we plan to explore the effects of choice of loss function in future work. 

One could also envision tasking the (D)CNN with learning the uncertainty directly\textemdash i.e. learning a mean and standard deviation for each pixel or even the full covariances\textemdash as opposed to estimating the uncertainty with an ensemble of models each making many predictions on noisy data. Note that while this would save significant computational expense the most reliable uncertainty estimate will always come from an ensemble approach due to the inherent nature of the local minima found via gradient descent each time the model is trained.

While the results shown here are promising, this paper is simply a proof of concept of this method and we believe many refinements to this methodology can and should be made before applying to real data.
For example, the ``blind'' test (D)CNN inversion shown in Figure \ref{fig:2Dtest} highlights issues that may arise in applying the current version of the (D)CNN method to more realistic data, highlighting the importance of de-trending.
The ``blind test'' performed here is also just one possible ``realistic'' BLR, thus in the future we will seek to benchmark our models with a variety of more realistic simulations in addition to this case.
Regardless of which more realistic simulations we use, a key improvement in future work will be designing a training set that includes more realistic observational errors and seasonal gaps,
as currently our training set includes only simple approximations of these effects. Ideally this would be done in such a way that is informed by the
expected data quality and observational cadence of whatever instrument was used to collect the data we wish to apply this method to,
thus in the future to obtain the best (D)CNNs may require training for each particular instrument and survey campaign.

Figure \ref{fig:similarityQSO} hints that in addition to the previously discussed architectural changes and improvements altering the training data-sets to be more similar to the expected predicted inputs could also improve model performance. 
This is another critically important future goal, as of course better and more realistic training sets will lead to better performance on real data. In the future we will seek to expand the plausible space of BLR models included in our training set, most urgently including models with
more realistic physics and radiative transfer such that our models can learn to predict the amplitude of the response as well as the shape, thus constraining properties of the gas itself in addition to just the geometry and kinematics of the BLR.
With large amounts of astronomical time series data expected in the near future from survey campaigns from Roman \citep{ROMAN_overview} and LSST \citep{LSST_manual} it is important that our analysis methods be able to keep up with this deluge of data. While the approach demonstrated here is not a panacea, as transfer learning is still required for each unique continuum lightcurve, this approach may help to ease the analysis burden for the interpretation of many future RM observations. We also plan to explore an even more general version of the (D)CNN in the near future, where instead of learning to each continuum instead the method could take as inputs a continuum and set of emission line lightcurves to generate predictions for a wide variety of sources even faster.

Obtaining the true 2D transfer function for a given source in this way is (physically) model agnostic, and after the true 2D $\Psi$ is recovered physical models can be compared with the resulting outputs to better constrain the fundamental nature of the BLR. There are many possible ways to match 1D observational products such as line profiles, delay profiles, and 1D response functions, and this degeneracy can only be fully broken by directly fitting the 2D velocity-delay maps such as those shown here. There are a growing number of real 2D RM datasets with the quality and cadence required to use this new method to recover transfer functions for real sources \citep{STORM_MODEL_SPECTRA,STORM2_II,RM3c27319,Wang_2025_SouthKoreaRM}, which we can then attempt to match with more physical models of the BLR to better understand its fundamental nature. 
In the future we hope to apply this method to these datasets as a natural next step, as well as more robustly compare the strengths and weaknesses of various inversion methods to better understand what regimes each method should be employed in.

\section{Acknowledgments}
This work was supported in part by NSF grants AST-1909711 and AST-2307983, and by an Alfred P. Sloan Research Fellowship (JD). This work utilized the Alpine high performance computing resource at the University of Colorado Boulder. Alpine is jointly funded by the University of Colorado Boulder, the University of Colorado Anschutz, and Colorado State University and with support from NSF grants OAC-2201538 and OAC-2322260. KL is especially grateful to Sajal Gupta for many helpful discussions over the course of the project. While the full ensembles of models are too large to host cheaply on the internet, sample models and code to reproduce the results in this work are available on the corresponding author's \href{https://github.com/kirklong/DCNN}{GitHub}, with the full ensembles and training/validation datasets available on request.
We are grateful to the anonymous referee for their thorough report, which greatly improved the quality and clarity of this paper, particularly in regards to quantifying this new method's performance.

\software{Julia, including:
          \texttt{BroadLineRegions.jl},
          \texttt{Flux.jl},
          \texttt{CairoMakie.jl},
          \texttt{Plots.jl},
          \texttt{MEMEcho}}

\newpage

\bibliography{citations}{}
\bibliographystyle{aasjournal}

\appendix 
\label{appendix}
\vspace{-6mm} 
\section{(D)CNN model architecture summary}\label{sec:model}
\vspace{-6mm}
\begin{figure*}[!htb]
    \centering
    \includegraphics[width=\textwidth,keepaspectratio]{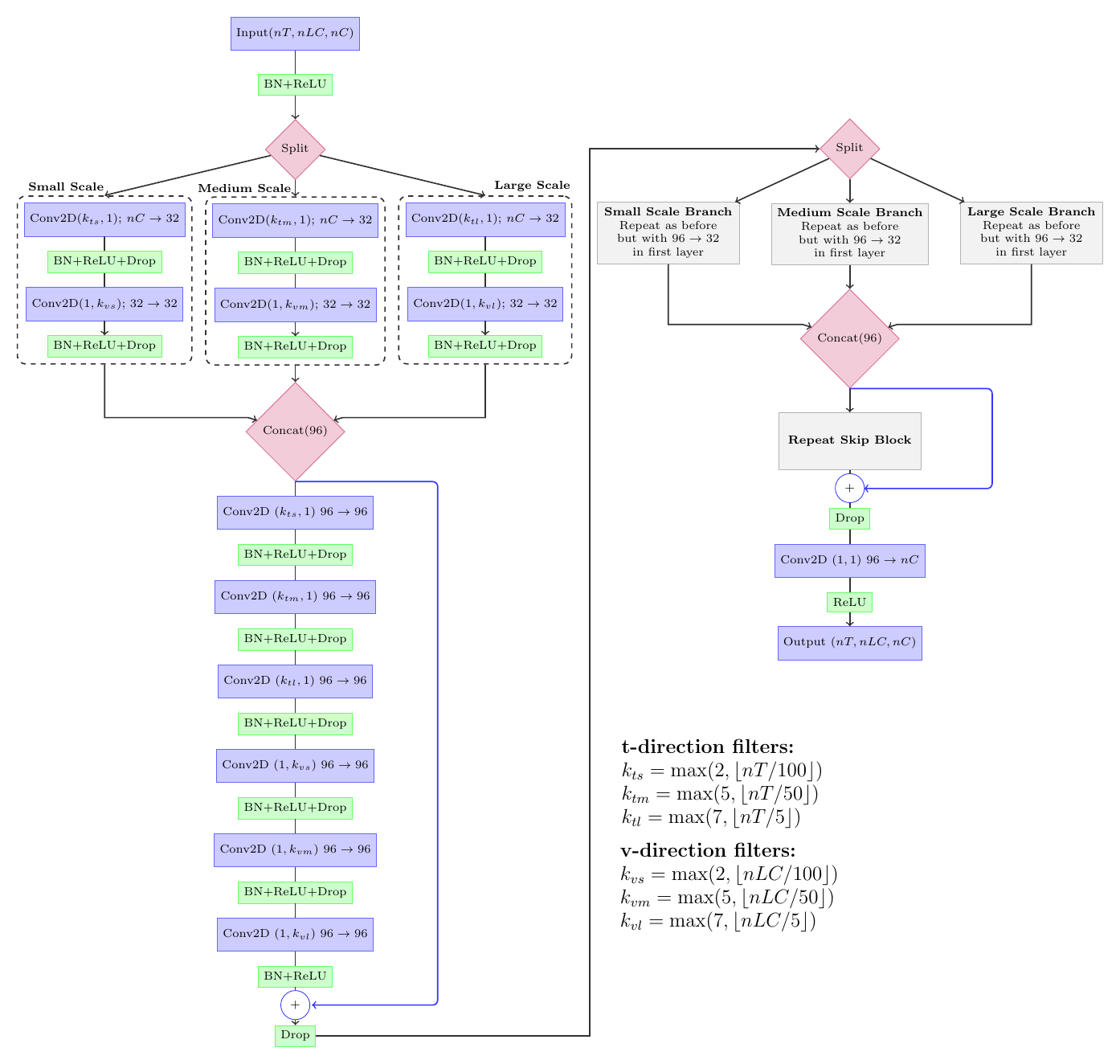}
    \caption{A schematic of the general architecture of our (D)CNN. The indigo boxes represent the input, output, and convolutional layers, with each convolutional layer including a description of the number of channels at that step (we fix $nC = 1$ in this work). The green boxes represent regularization\textemdash BatchNorm (BN) and Dropout (Drop)\textemdash and activation\textemdash ReLU\textemdash layers that occur after each convolutional layer. The red diamonds represent split/join operations, where in the case of joining the number of channels after joining is denoted in parentheses. The left panel shows one split/join and skip connection chain in full, while the right panel showcases that our model architecture essentially performs these operations twice. Note that while the full 2D schematic is shown in the 1D case the general pattern is the same but the convolutions are 1D and only along the time dimension thus there are $\sim$ half the number of layers than in the 2D case. When the number of velocity channels is increased the gradients become large in the 2D case, and thus sometimes memory requirements necessitate using only one split/concat loop instead of both.}
    \label{fig:model}
\end{figure*}
A flowchart for the (D)CNN model architecture is shown in Figure \ref{fig:model}, as described in the main text in Section \ref{sec:models}.

\section{(D)CNN hyperparameters}\label{sec:hyperparams}
Table \ref{tab:hyperparams} includes a summary of all of the hyperparameters ued in training our (D)CNN models.
\startlongtable
\begin{deluxetable*}{lll}
\tablecaption{Summary of (D)CNN hyperparameters used in this work. Hyperparameters marked ``Tuned'' were determined experimentally with a variety of possibilities tried in testing,
while all others were fixed based on standards in the literature.\label{tab:hyperparams}}
\tablecolumns{3}
\tablehead{Hyperparameter & Value & Fixed/Tuned}
\startdata
\cutinhead{Architecture}
Multi-scale branches & 3 (small, medium, large) & Tuned \\
Small time filter ($k_{ts}$) & $\max(2,\,\lfloor n_T/100\rfloor)$ & Tuned \\
Medium time filter ($k_{tm}$) & $\max(5,\,\lfloor n_T/50\rfloor)$ & Tuned \\
Large time filter ($k_{tl}$) & $\max(7,\,\lfloor n_T/5\rfloor)$ & Tuned \\
Velocity filters ($k_{vs}$, $k_{vm}$, $k_{vl}$) & Analogous formulas with $n_\text{LC}$ & Tuned \\
Channels per branch & 32 & Tuned \\
Activation function & ReLU & Fixed \\
Regularization & BatchNorm + Dropout (rate\,=\,0.1) & Fixed \\
Weight initialization & Glorot normal & Fixed \\
Padding & Same (output dims = input dims) & Fixed \\
Skip connections & Additive (ResNet-style) & Fixed \\
Inception+residual blocks & 2 (1D), 1 (2D) & Tuned \\
\cutinhead{Training}
Optimizer & Adam ($\beta_1{=}0.9$, $\beta_2{=}0.999$, $\epsilon{=}10^{-8}$) & Fixed \\
Learning rate & $1.5\times10^{-4}$ & Tuned \\
Loss function & MSE & Fixed \\
Batch size & 64 & Tuned \\
Max.\ epochs & 100 (typical $\sim$50 before autostop) & Fixed \\
Ensemble size & 100 & Fixed \\
Train/validation split & 70\%/30\% & Fixed \\
Autostop window & 5 epochs & Tuned \\
Autostop threshold & $10^{-5}$ & Tuned \\
LR reduction on plateau & Halved (up to $2\times$) & Fixed \\
Post-training ensemble cut & $1\sigma$ from mean validation loss & Tuned \\
\enddata
\end{deluxetable*}
\section{1D transfer function parameter ranges}\label{sec:1Dparams}
\begin{figure}
    \centering
    \includegraphics[width=0.48\textwidth,keepaspectratio]{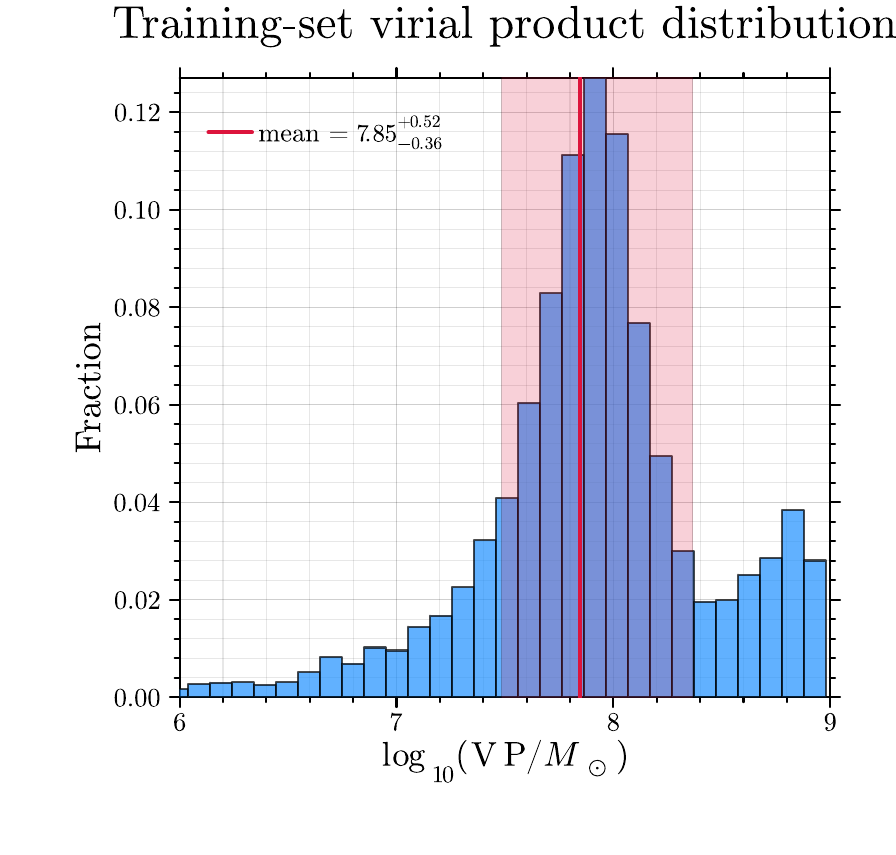}
    \caption{Distribution of virial products ($c\bar{\tau}\Delta v^2/G$) for all of our samples, as calculated from the mean $\Delta v^2$ of the velocity distribution and mean delay $\bar{\tau}$ for each sample.
    We show the virial product calculated this way as the ``shape''-based BLRs do not have intrinsic black hole masses or radii, and thus this approximates the distribution of ``masses'' used in training.}
    \label{fig:viralDistribution}
\end{figure}

The parameter ranges used to generate each class of 1D synthetic transfer function for training are listed in Table~\ref{tab:1Dparams}. All model types are generated over a peak delay range of $\tau_\text{peak}\in[5,25]$~days sampled uniformly. For BLR kinematic models (disk-wind and cloud), either the mean BLR radius $\bar{r}$ or the black hole mass factor $M_\text{fac}$ is sampled directly with equal probability; the other quantity is derived from the peak delay constraint. A histogram showing the assumed virial product distribution used in training (for both 1D and 2D cases) is shown in Figure \ref{fig:viralDistribution}.

\startlongtable
\begin{deluxetable*}{llll}
\tablecaption{Parameter ranges for each class of 1D synthetic transfer function used in training. The peak delay $\tau_\text{peak}\in[5,25]$~days is shared by all model types and sampled uniformly.\label{tab:1Dparams}}
\tablecolumns{4}
\tablehead{Model & Parameter & Range & Distribution}
\startdata
Exponential Decay & No additional free parameters & \nodata & \nodata \\
\hline
Gaussian & Width ($\sigma_t$) & 1--$\min(10,\,\tau_\text{peak})$~days & Uniform \\
\hline
Double-Gaussian & Width ($\sigma_t$) & 1--10~days & Uniform \\
 & Secondary peak ($\tau_2$) & $\tau_\text{peak}$--$3\,\tau_\text{peak}$ & Uniform \\
 & Secondary amplitude scale & 0--$5\,e^{-\tau_2/\tau_\text{peak}}$ & Uniform \\
\hline
Accretion Disk & Characteristic time ($\tau_0$) & 5--25~days & $=\tau_\text{peak}$ \\
 & Temperature index ($b$) & $3/4$ & Fixed \\
\hline
BLR Disk-Wind & Inclination ($i$) & $5^\circ$--$85^\circ$ & Uniform \\
 & Mean radius ($\bar{r}$) & 500--2000~$r_s$ & Uniform\tablenotemark{a} \\
 & Mass factor ($M_\text{fac}$) & 0.1--2.0~($\times10^8\,M_\odot$) & Uniform\tablenotemark{a} \\
 & Radius factor ($r_\text{fac}$) & 2--100 & Uniform \\
 & $f_1,\,f_2,\,f_3,\,f_4$ & 0--1 (each) & Uniform \\
 & $S_\alpha$ & 0--2 & Uniform \\
\hline
BLR Cloud & Inclination ($i$) & $5^\circ$--$85^\circ$ & Uniform \\
 & Mean radius ($\bar{r}$) & 500--2000~$r_s$ & Uniform\tablenotemark{a} \\
 & Mass factor ($M_\text{fac}$) & 0.1--2.0~($\times10^8\,M_\odot$) & Uniform\tablenotemark{a} \\
 & Anisotropy ($F$) & 0--1 & Uniform \\
 & $\kappa$ & $-0.5$--0.5 & Uniform \\
 & $\xi$ & 0--1 & Uniform \\
 & $\gamma$ & 0--5 & Uniform \\
 & Elliptical fraction ($f_\text{ell}$) & 0--1 & Uniform \\
 & Flow fraction ($f_\text{flow}$) & 0--1 & Uniform \\
 & Ellipse angle ($\theta_e$) & $0^\circ$--$90^\circ$ & Uniform \\
 & Turbulence ($\sigma_{\rho,r}$, $\sigma_{\rho,c}$, $\sigma_{\theta,r}$, $\sigma_{\theta,c}$, $\sigma_t$) & 0.1--1.0 (each) & Uniform \\
 & Radial power-law index ($\beta$) & 0--3 & Uniform \\
 & Opening angle ($\theta_o$) & $0^\circ$--$90^\circ$ & Uniform \\
 & Number of clouds ($N_\text{clouds}$) & 500{,}000 & Fixed \\
\enddata
\tablenotetext{a}{Either $\bar{r}$ or $M_\text{fac}$ is sampled directly (50\% probability each); the other is derived from the peak delay constraint. Given the complicated nature of this sampling, we include 
a plot showing the resulting distribution of virial products used in training in Figure \ref{fig:viralDistribution}.}
\end{deluxetable*}

\newcommand{\twodnotecell}[1]{\parbox[t]{5cm}{#1}}
\section{Sample 2D transfer functions and trailed spectrograms as well as 2D transfer function parameter ranges}\label{sec:sample2D}
\begin{figure*}[!htb]
    \centering
    \includegraphics[width=0.87\textwidth,keepaspectratio]{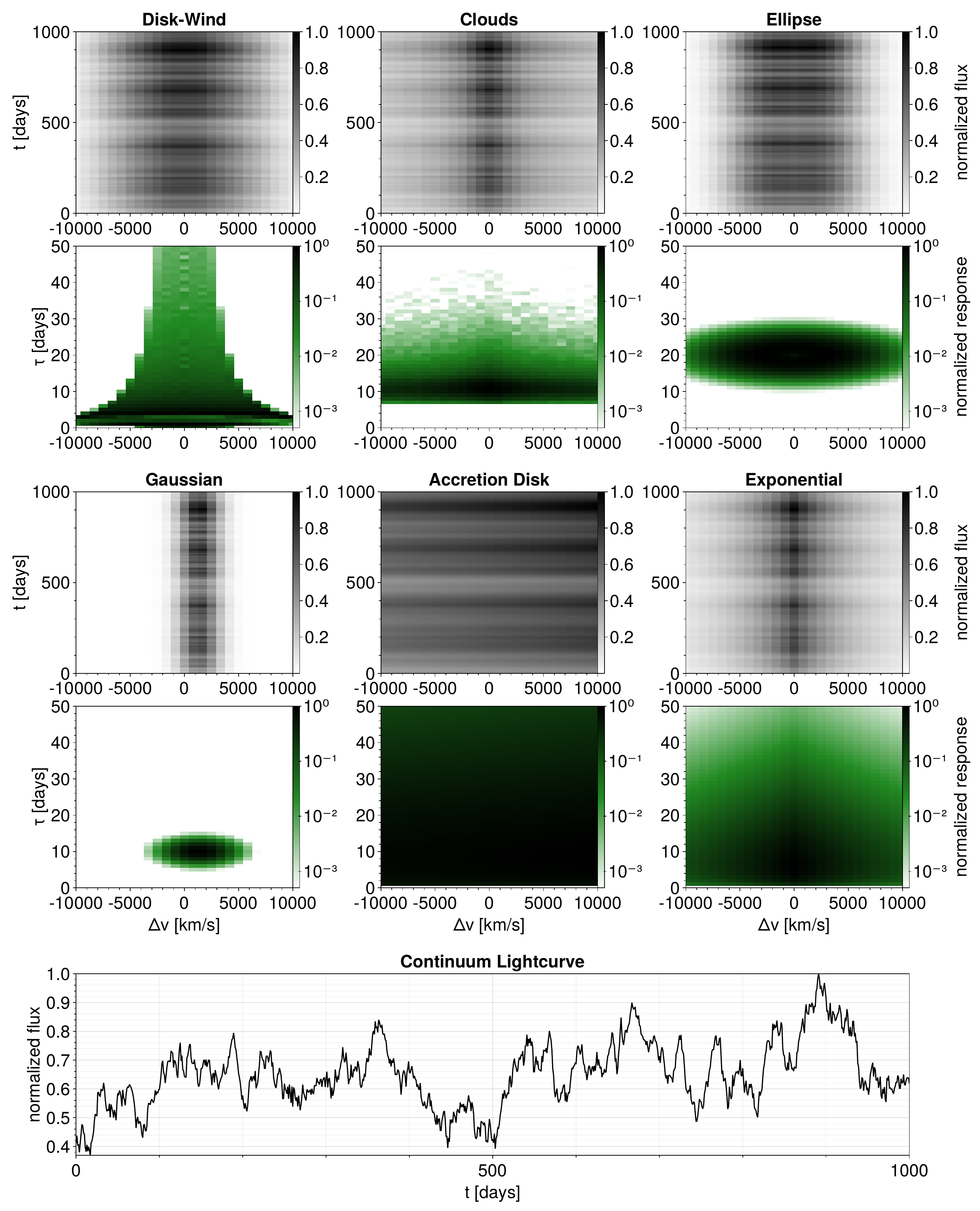}
    \caption{Similar to Figure \ref{fig:1DSampleResults}, here we show representative samples from each class of ``basis'' transfer function used in training our models. \textbf{Top grid:} representative trailed spectrograms and transfer functions for each basis function. The bottom panel of each grid entry shows the 2D transfer function and the top panel shows the spectrogram (2D visualization of all of the individual lightcurves) that results after convolving with the continuum lightcurve and adding errors. The top panels thus present sample inputs to the (D)CNN model and the bottom panel the expected outputs. \textbf{Bottom:} the continuum lightcurve used in these generating these examples.}
    \label{fig:2DSamples}
\end{figure*}

The parameter ranges for the 2D velocity-delay transfer functions used in training are listed in Table~\ref{tab:2Dparams}. Parameters shared with the 1D models (Table~\ref{tab:1Dparams}) are not repeated\textemdash only parameters that are unique to the 2D case or that differ from the 1D values are listed. Figure \ref{fig:2DSamples} showcases 2D example transfer functions for each class of sample included in training.

\startlongtable
\begin{deluxetable*}{llll}
\label{sec:2Dparams}
\tablecaption{Parameter ranges for 2D synthetic transfer functions. Only parameters unique to the 2D case or that differ from the 1D values (Table~\ref{tab:1Dparams}) are listed.\label{tab:2Dparams}}
\tablecolumns{4}
\tablehead{Model & Parameter & Range (2D) & Notes vs.\ 1D}
\startdata
Exponential Decay & Velocity scale ($\Psi_{v,\max}$) & $U(\lvert v_\text{min}/10\rvert,\;\lvert 3\,v_\text{max}/4\rvert)$ & \twodnotecell{Velocity profile $\propto e^{-|v|/\Psi_{v,\max}}$} \\
\hline
Gaussian & Velocity scale ($\Psi_{v,\max}$) & $U(\lvert v_\text{min}/10\rvert,\;\lvert 3\,v_\text{max}/4\rvert)$ & \twodnotecell{Center of Gaussian in velocity} \\
 & Velocity width ($\sigma_v$) & 100--$\min(2000,\,\Psi_{v,\max})$~km\,s$^{-1}$ & \twodnotecell{Capped by $\Psi_{v,\max}$} \\
\hline
Elliptical Ring & Time width ($\sigma_t$) & 1--$\min(10,\,\tau_\text{peak})$~days & \twodnotecell{Now an actual ring in $(\tau, v)$ as opposed to approximating two Gaussians at different times} \\
 & Velocity width & $U(\lvert v_\text{min}/10\rvert,\;\lvert v_\text{max}/2\rvert)$ & \twodnotecell{New parameter} \\
 & Ellipse width & 0.5--5.0 & \twodnotecell{New parameter} \\
 & Time center ($\tau_c$) & $\tau_\text{peak}$--$2\,\tau_\text{peak}$ & \twodnotecell{New parameter} \\
\hline
Accretion Disk & Velocity dependence & $\lambda = 1/(1-v/c)$ & \twodnotecell{Doppler shifted per velocity bin} \\
\hline
BLR Disk-Wind & Mean radius ($\bar{r}$) & 500--1200~$r_s$ & \twodnotecell{Slightly narrower than 1D to focus model learning at shorter times}\\
\hline
BLR Cloud & (All parameters same as 1D) & \nodata & \twodnotecell{Velocity structure intrinsic to model} \\
\enddata
\end{deluxetable*}

\section{Analytic face-on accretion disk prescription}\label{sec:KDM}
For a face-on accretion disk with a power-law temperature profile, $T \propto r^{-b}$, the delay distribution is given by the following proportionality:

\begin{equation}
\Psi_\nu(\tau|\lambda) \propto \left(\frac{\lambda_0}{\lambda}\right)^2\left(\frac{\tau}{\tau_0}\right)^{3b-2} W(x)
\end{equation}

With $x = (\lambda_0/\lambda)(\tau/\tau_0)^b$ and the dimensionless factor $W(x) = \frac{x^2/2}{\cosh x -1}$. We assume a steady-state disk model, which gives $b=3/4$ and $3b-2=1/4$. We keep the peak wavelength $\lambda_0$ fixed, and alter the characteristic time $\tau_0$ in generating our models of this kind, with each being normalized independently such that we do not specify an exact temperature or other physical quantities for the disk but instead only keep the proportionalities above consistent.
\section{(D)CNN systematic bias analysis}\label{sec:bias}
To quantify whether the (D)CNN models exhibit systematic directional bias rather than purely random errors, we compute flux-weighted first and second moments (centroid and RMS width) of both the predicted and ground-truth transfer functions across the full test sets.
For the 1D models ($N_{\rm{test}}\sim 4\times 10^4$) we measure the centroid of the lag $\bar{\tau}$ and the lag width $\sigma_\tau$,
while for the 2D models ($N_{\rm{test}}\sim 10^4$) we additionally measure the centroid of the velocity $\bar{v}$ and the velocity width $\sigma_v$.
We can then compare the per-sample bias ($\Delta = \text{predicted} - \text{truth}$) and investigate if there are any systematic trends in the distribution of biases with
respect to these four quantities.

\begin{figure*}[!htb]
    \centering
    \includegraphics[width=0.87\textwidth,keepaspectratio]{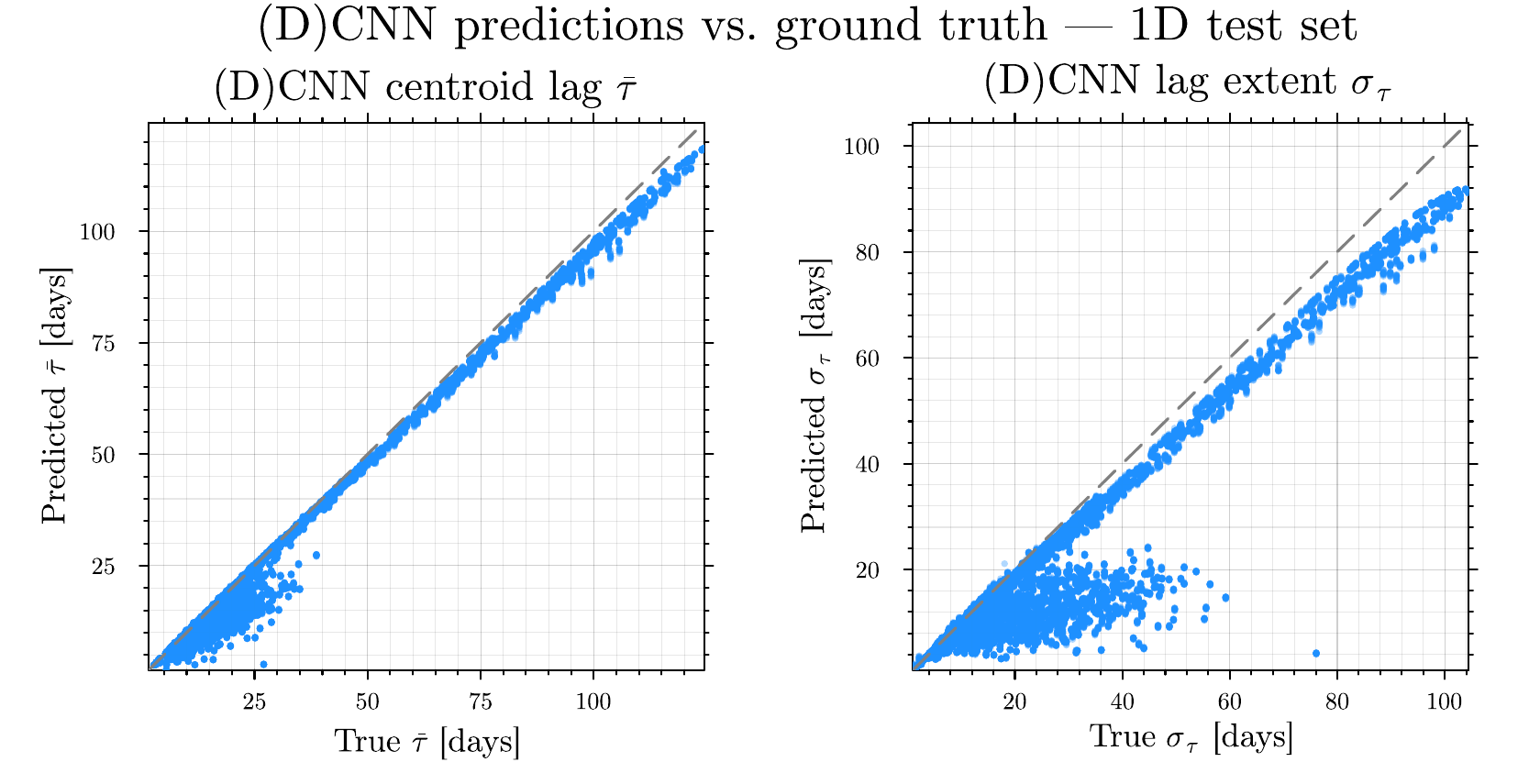}
    \caption{Predicted vs.\ ground-truth moment statistics for the 1D test set ($N_{\rm{test}}\sim 4\times 10^4$ samples). Left: flux-weighted centroid lag $\bar{\tau}$. Right: RMS lag width $\sigma_\tau$. The dashed line shows the 1:1 relation.}
    \label{fig:bias1Dscatter}
\end{figure*}

\begin{figure*}[!htb]
    \centering
    \includegraphics[width=0.87\textwidth,keepaspectratio]{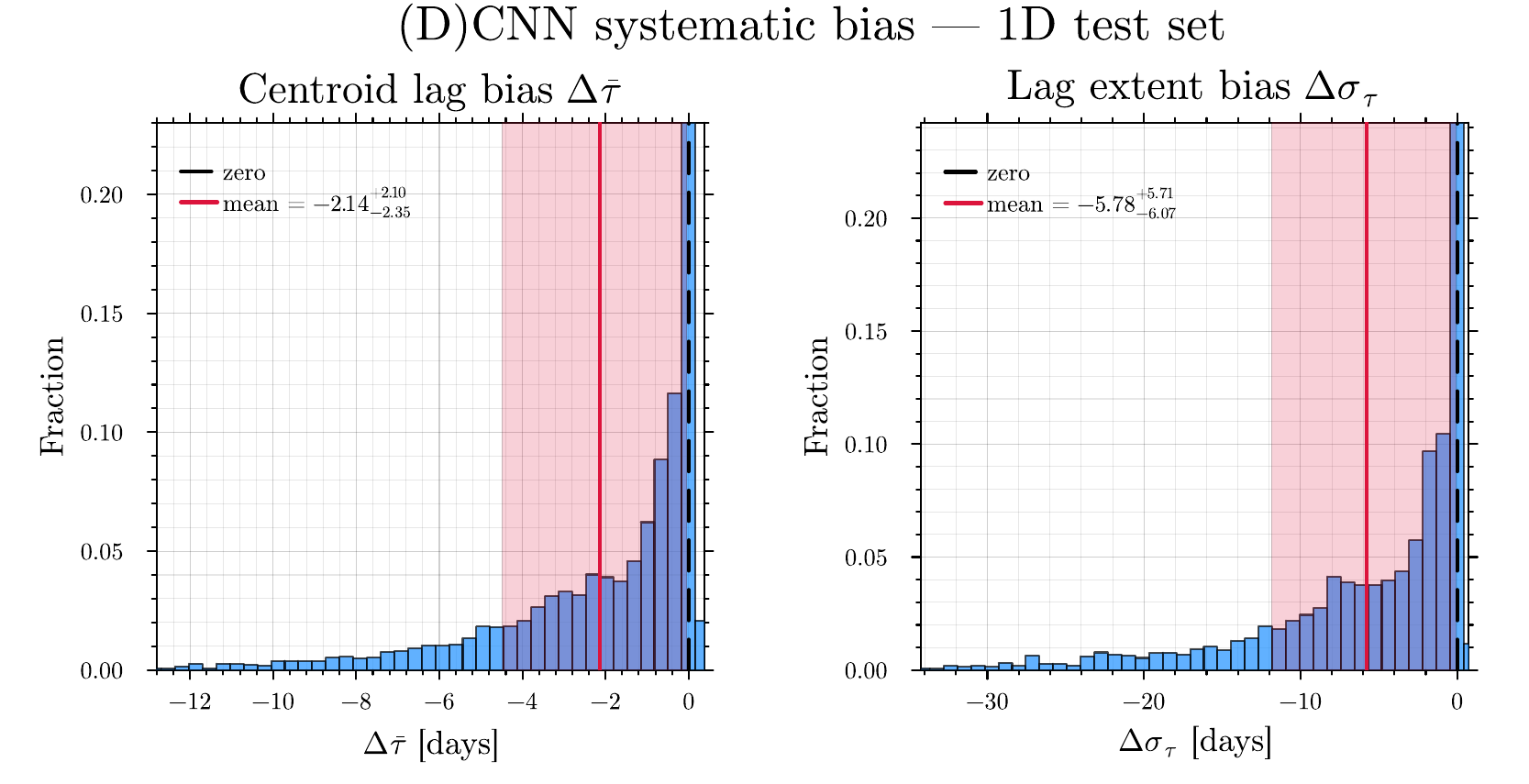}
    \caption{Distributions of per-sample bias $\Delta = \text{predicted} - \text{truth}$ for the 1D test set. Left: centroid lag bias $\Delta\bar{\tau}$. Right: lag width bias $\Delta\sigma_\tau$. The vertical dashed line marks zero; the solid line marks the mean bias with asymmetric 16th--84th percentile uncertainties as labeled. The shaded region spans the 16th--84th percentile of the bias distribution. Note that the time axis is in units of array elements; the underlying physical timescale is set by the cadence of the input lightcurve.}
    \label{fig:bias1Dhist}
\end{figure*}

\begin{figure*}[!htb]
    \centering
    \includegraphics[width=0.87\textwidth,keepaspectratio]{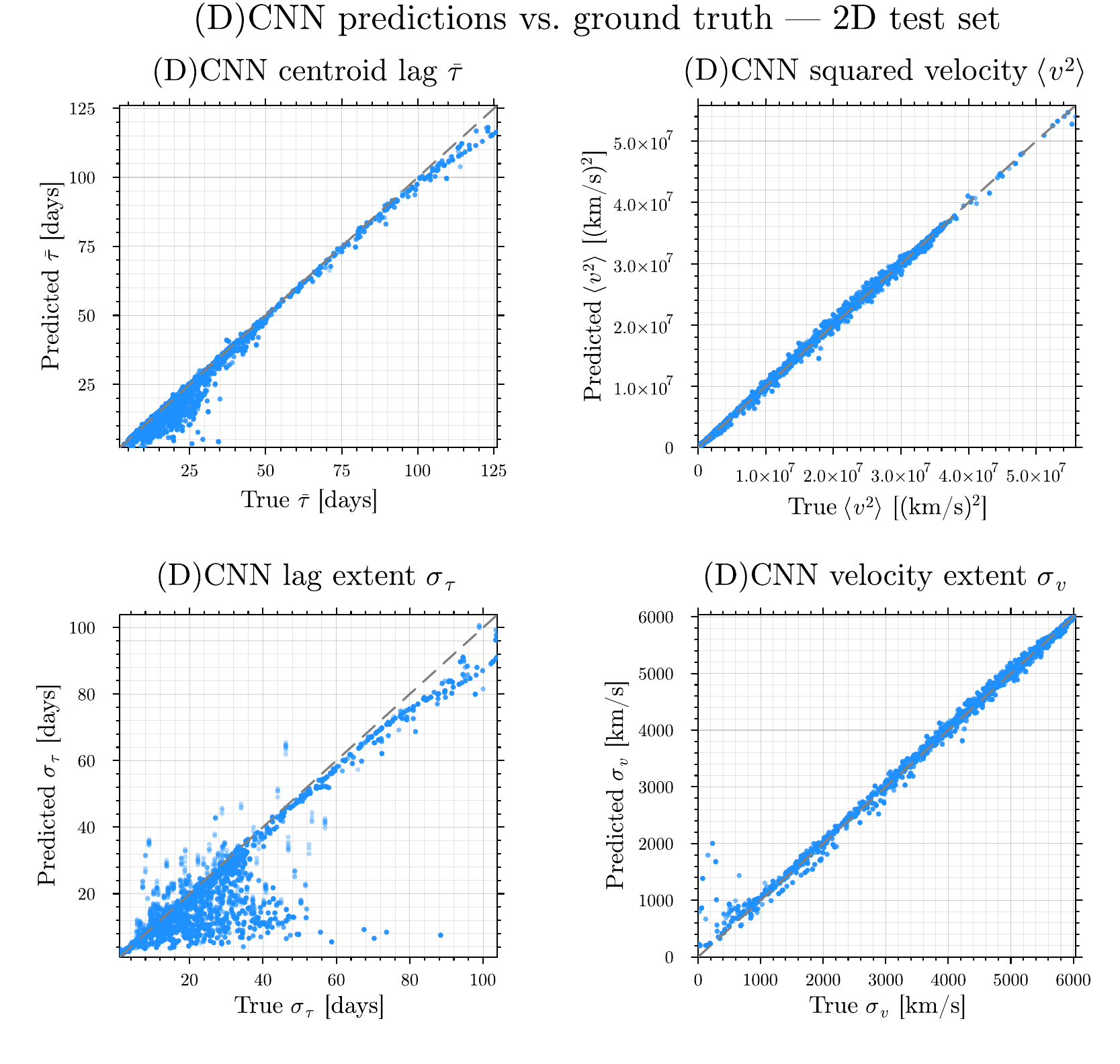}
    \caption{Predicted vs.\ ground-truth moment statistics for the 2D test set ($N_{\rm{test}}\sim 10^4$ samples). Top row: centroid lag $\bar{\tau}$ and centroid of mean squared velocity $\langle v^2\rangle$\textemdash we show the
    mean squared velocity centroid instead of just the unsquared velocity centroid as many of the training set samples have an average velocity of zero but show distinct structures away from zero velocity. Bottom row: RMS lag width $\sigma_\tau$ and velocity width $\sigma_v$. The dashed line shows the 1:1 relation.
    Note that in combining these panels one has a sense for how the model recovers the virial product of the system, which it thus does quite well for systems with delays shorter than the maximum.}
    \label{fig:bias2Dscatter}
\end{figure*}

\begin{figure*}[!htb]
    \centering
    \includegraphics[width=0.87\textwidth,keepaspectratio]{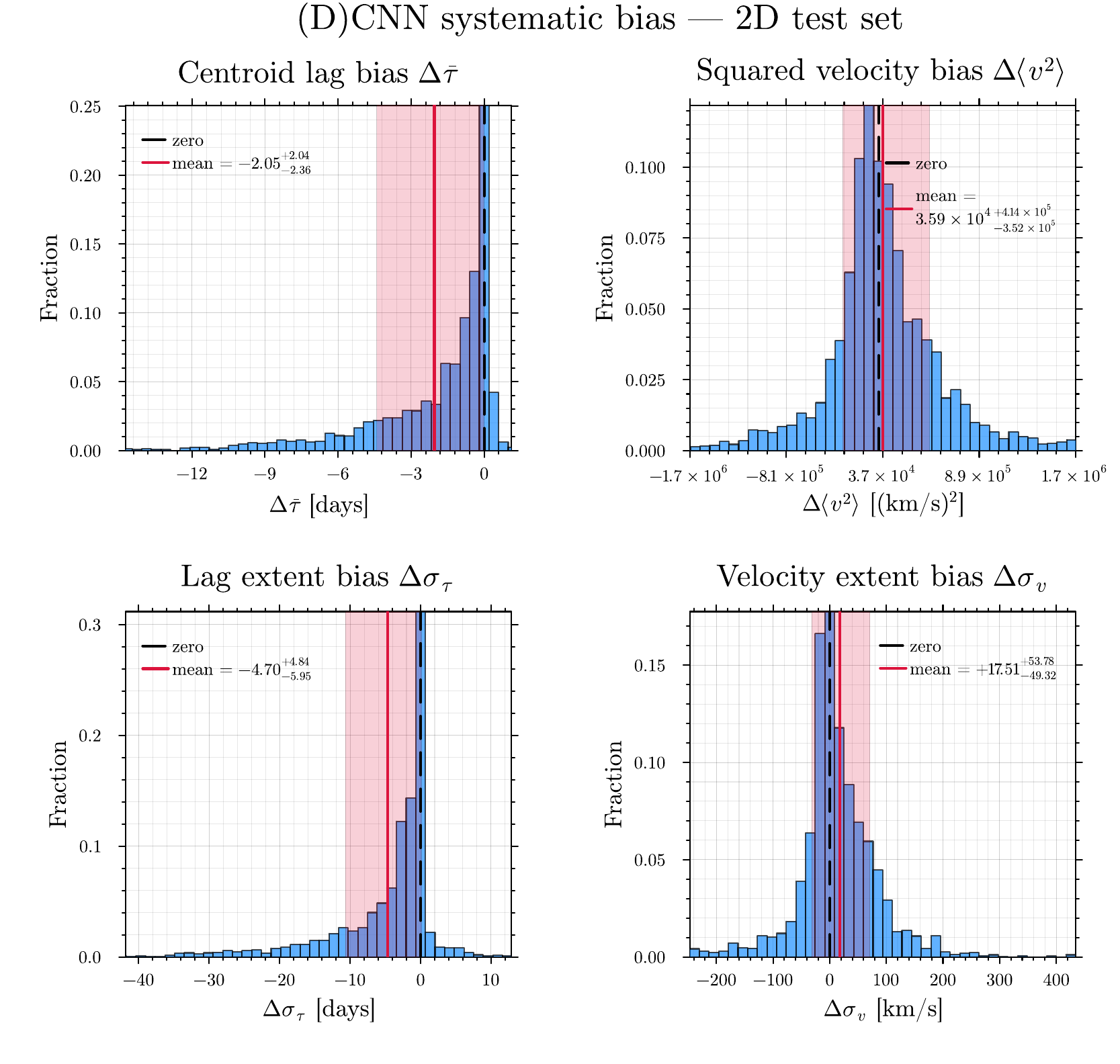}
    \caption{Distributions of per-sample bias $\Delta = \text{predicted} - \text{truth}$ for the 2D test set. Top row: centroid lag bias $\Delta\bar{\tau}$ and centroid velocity bias $\Delta\bar{v}$. Bottom row: lag width bias $\Delta\sigma_\tau$ and velocity width bias $\Delta\sigma_v$. Axes are clipped to the 0.5th--99.5th percentile to suppress extreme outliers. Annotations and shading follow the same convention as Figure~\ref{fig:bias1Dhist}.}
    \label{fig:bias2Dhist}
\end{figure*}

Figures~\ref{fig:bias1Dscatter} and \ref{fig:bias1Dhist} show these results for the 1D cases (using the missing data test set and corresponding ensemble of models),
while Figures~\ref{fig:bias2Dscatter} and \ref{fig:bias2Dhist} show the 2D results.
In all cases the models exhibit a systematic trend towards underrepresenting the ground-truth delay values in both centroid and width,
meaning predictions are systematically shifted toward shorter delays and narrower profiles.
The velocity quantities show comparatively smaller fractional biases, indicating relatively random noise about the ground truth. This is
consistent with our visual observations, as we often say that in velocity space the models would have the right morphology but could be off by a $\sim$ pixel
in width in velocity space, and the models struggle in general with longer delays, an error that is physically motivated by the nature of the problem.
The shaded ribbons in the histogram panels span the 16th--84th percentile range of each bias distribution,
with the asymmetric uncertainties on the mean derived from these same percentiles.

Note that while the labels on the axes are in units of days and km/s, really both the delay and velocity axes are simply in units of array elements,
where here we have shown them
in days and km/s as this the scale we used for our models in the training set, but in principle both could be rescaled. The models used in training
were generated with a cadence of 1 day (such that shifting 1 array element = shifting 1 day) and a velocity binning of 400 km/s
(such that shifting 1 array element = shifting 400 km/s). Thus the mean offset of $\sim$ 2 days corresponds to the models on average being wrong
by about 2 array elements in the delay dimension, while the mean offset of just $\sim$ -1 km/s in the velocity dimensions indicates that on average
the models are not systematically shifted in velocity space and instead the errors are more random in nature (i.e. shifted by a bin either positively or negatively
with equal probability).

Note that from the scatter plots (Figures~\ref{fig:bias1Dscatter} and \ref{fig:bias2Dscatter}) it is clear that the models almost \textit{never} overpredicts
the mean delay, and they consistently struggle with the longest delay values as expected given that the input lightcurves are limited to $\sim$ 3 years in duration in this scaling.
Thus this is likely not a limitation of this method but instead a limitation of the training set, as the model has not been shown enough information to robustly cover times at longer delays
given the length of the input lightcurves.
The turn off point could be used in future studies to motivate
the length of the training set in time space, i.e. if here our models struggle to predict delays longer than $\sim$ 1/10th of the total time shown in
training this could likely be mitigated in the future by simply training on longer time series, with the length set by the maximum delay one
wishes to most accurately predict. Previous analytic calculations for reverberation mapping campaigns suggest that the duration of observations must exceed
the desired timescale by a factor of at least three \citep{Horne2004_RM_requirements}, thus our model performance starting to degrade over $\sim$ 1/10th of the input lightcurve duration is a bit smaller
yet consistent with this idea. Our models are not graded on their ability to recover the mean delay, however, but instead the gradients will be largest where the signal is largest which is usually at
shorter delays, but if in the future we wanted to better approach the theoretical limit we could include an explicit penalty in the loss function for wrongly predicting the mean delay.

There is also larger spread at lower delays, and in particular there appears to be a second population that is slightly farther off the 1:1 line than the rest of the samples.
This subpopulation is almost entirely composed of the hardest classes of samples the model was exposed to\textemdash physical BLRs + combined, multi-component BLRs\textemdash confirming that the most
complicated structures are the hardest to accurately recover. Also
note that the spread of predicted widths is larger than the spread of predicted centroids, which is physically motivated
as well as the width is more difficult to predict than the centroid. Interestingly it appears that these biases are correlated with each other,
such that when a model underpredicts the centroid it also tends to underpredict the width. From these panels it is clear that future iterations of
this methodology should be aware of this potential bias in the delays, and in particular design training sets that are sensitive to this issue, while the velocity predictions appear to be relatively unbiased in any systematic
way by comparison.

\end{document}